\documentclass[letterpaper,twocolumn,10pt]{article}
\usepackage{usenix}
\usepackage[mathscr]{eucal}
\usepackage{tikz}
\usepackage{amsmath}
\usepackage{shadowtext}
\usepackage{filecontents}
\shadowoffset{2pt}

\usepackage{amsmath,amsfonts,bm}

\def\eqref#1{equation~\ref{#1}}

\def\1{\bm{1}}

\def\rvy{{\mathbf{y}}}

\DeclareMathAlphabet{\mathsfit}{\encodingdefault}{\sfdefault}{m}{sl}
\SetMathAlphabet{\mathsfit}{bold}{\encodingdefault}{\sfdefault}{bx}{n}

\def\sD{{\mathbb{D}}}

\def\sI{{\mathbb{I}}}

\newcommand{\final}[1]{\textcolor{black}{#1}}
\newcommand{\revise}[1]{\textcolor{black}{#1}}
\newcommand{\ie}{\textit{i}.\textit{e}.}
\newcommand{\eg}{\textit{e}.\textit{g}.}
\newcommand{\etal}{\textit{et} \textit{al}.}
\newcommand{\etc}{\textit{etc}.}
\newcommand{\cf}{\textit{c}.\textit{f}. }
\newcommand{\method}{$\mathcal{X}$-\emph{Adv} }

\usepackage{pifont}
\usepackage[perpage,symbol*]{footmisc}
\DefineFNsymbols{circled}{{\ding{192}}{\ding{193}}{\ding{194}}
{\ding{195}}{\ding{196}}{\ding{197}}{\ding{198}}{\ding{199}}{\ding{200}}{\ding{201}}}
\setfnsymbol{circled}

\newcommand\myfootnotestyle[1]{\ifcase#1 \or \ding{182}\or \ding{183}\or
\ding{184}\or \ding{185}\or \ding{186}\or \ding{187}%
\or \ding{188}\or \ding{189}\or \ding{190}\or \ding{191}\else *\fi\relax}

\begin{document}

\date{}

\title{\shadowtext{\textcolor{black}{$\mathcal{X}$}}-\emph{Adv}: Physical Adversarial Object Attacks against X-ray \\Prohibited Item Detection}

\author{
{\rm Aishan Liu$^1$*, Jun Guo$^1$*, Jiakai Wang$^2$, Siyuan Liang$^3$, Renshuai Tao$^1$, } \\
{\rm Wenbo Zhou$^4$, Cong Liu$^5$, Xianglong Liu$^{1,2,6\dag}$, Dacheng Tao$^7$}
\\ $^1$Beihang University, $^2$Zhongguancun Laboratory, $^3$Chinese Academy of Sciences, \\ $^4$University of Science and Technology of China, $^5$iFLYTEK, \\ $^6$Hefei Comprehensive National Science Center, $^7$JD Explore Academy
}

\maketitle

\footnotetext{* Equal contribution.}
\footnotetext{$^\dag$ Corresponding author.}

\begin{abstract}
Adversarial attacks are valuable for evaluating the robustness of deep learning models. Existing attacks are primarily conducted on the visible light spectrum (\eg, pixel-wise texture perturbation). However, attacks targeting texture-free X-ray images remain underexplored, despite the widespread application of X-ray imaging in safety-critical scenarios such as the X-ray detection of prohibited items. In this paper, we take the first step toward the study of adversarial attacks targeted at X-ray prohibited item detection, and reveal the serious threats posed by such attacks in this safety-critical scenario. Specifically, we posit that successful physical adversarial attacks in this scenario should be specially designed to circumvent the challenges posed by color/texture fading and complex overlapping. To this end, we propose \method to generate physically printable metals that act as an adversarial agent capable of deceiving X-ray detectors when placed in luggage. To resolve the issues associated with color/texture fading, we develop a differentiable converter that facilitates the generation of 3D-printable objects with adversarial shapes, using the gradients of a surrogate model rather than directly generating adversarial textures. To place the printed 3D adversarial objects in luggage with complex overlapped instances, we design a policy-based reinforcement learning strategy to find locations eliciting strong attack performance in worst-case scenarios whereby the prohibited items are heavily occluded by other items. To verify the effectiveness of the proposed $\mathcal{X}$-\emph{Adv}, we conduct extensive experiments in both the digital and the physical world (employing a commercial X-ray security inspection system for the latter case). Furthermore, we \final{present the physical-world X-ray adversarial attack dataset XAD}. We hope this paper will draw more attention to the potential threats targeting safety-critical scenarios. Our codes and XAD dataset are available at \url{https://github.com/DIG-Beihang/X-adv}.
\end{abstract}

\section{Introduction}

Deep neural networks (DNNs) have achieved remarkable performance across a wide area of applications \cite{Krizhevsky2012ImageNet,bahdanau2014neural,Hinton2012Deep}. Recently, deep learning has been introduced into safety-critical scenarios such as X-ray security inspection in public transportation hubs (\eg, airports). In this scenario \cite{wei2020occluded,Tao:ICCV21,Tao:CVPR22,miao2019sixray}, deep-learning-based detectors are utilized to assist inspectors in identifying both the presence and location of prohibited items (\eg, pistols and knives) during X-ray scanning. This approach significantly reduces the amount of human labor required and helps to protect the public from severe risks.

\begin{figure}[!t]
\vspace{-0.1in}
	\begin{center}
		\includegraphics[width=1.05\linewidth]{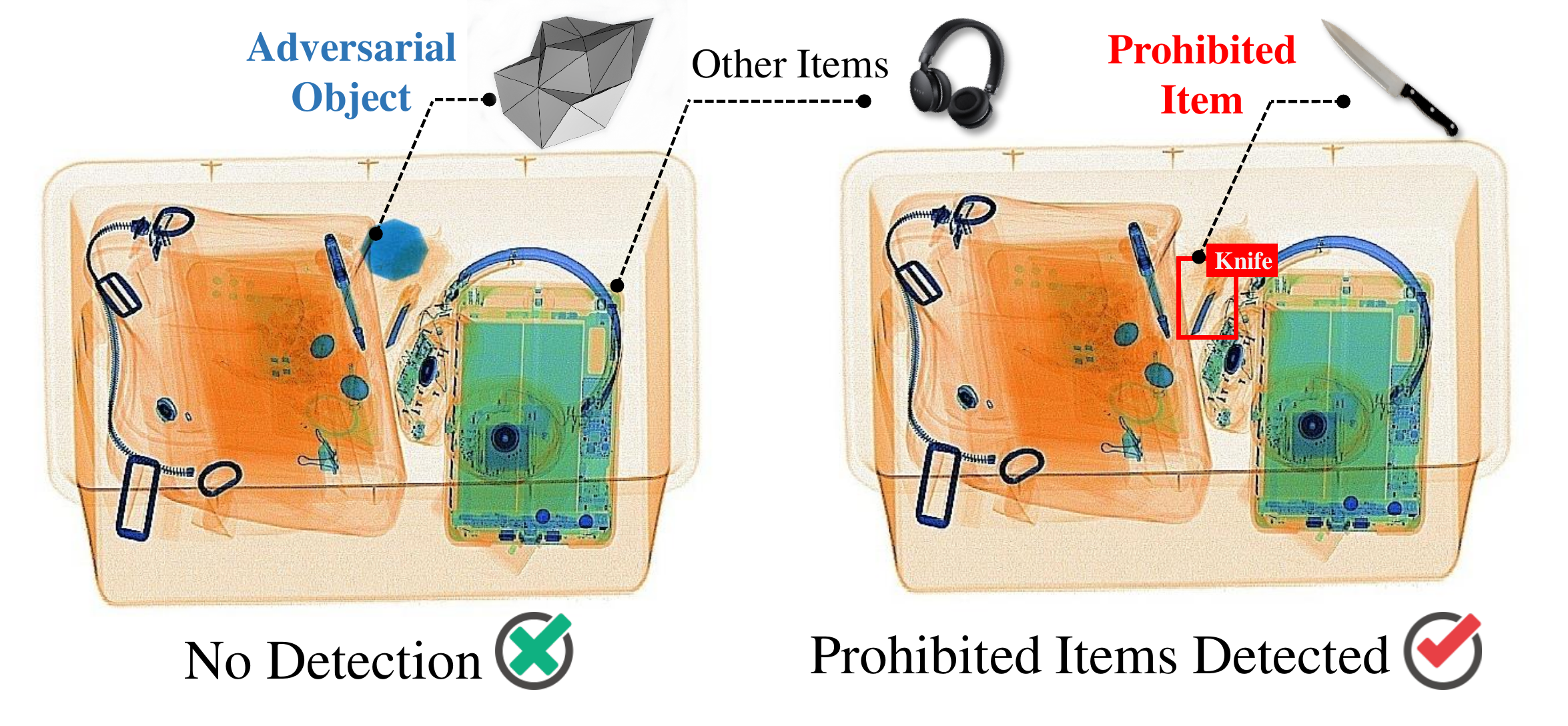}
	\end{center}
	\vspace{-0.2in}
	\caption{Illustration of physical-world adversarial attacks on X-ray security inspection. This paper proposes \method to generate physically realizable 3D adversarial objects. During X-ray scanning, the detector can detect prohibited items in the right image, but our adversarial objects deceive the detector into failing to detect prohibited items in the left image.}
	\label{fig:frontpage}
	\vspace{-0.1in}
\end{figure}

Despite their promising performance, DNNs are vulnerable to \emph{adversarial examples} \cite{szegedy2013intriguing}. These elaborately designed perturbations are imperceptible to human vision, but can easily mislead DNNs {into} making wrong predictions, thus threatening practical deep learning applications \cite{kurakin2016adversarial,Liu2019Perceptual,Liu2020Spatiotemporal}. By contrast, adversarial examples
are also beneficial for evaluating and better understanding the robustness of DNNs \cite{zhang2020interpreting,zhang2019interpreting,xie2020adversarial,liu2022harness,tang2021robustart}. In the past years, extensive research has been conducted into performing adversarial attacks on natural images (visual light); however, the robustness of texture-free X-ray images (such as in the context of X-ray prohibited item detection) remains underexplored. This sparsity of research presents a severe risk to the safety of the general public, as it increases their vulnerability to attack.

In this paper, we take the first step in physical-world adversarial attacks on the X-ray prohibited item detection scenario, \revise{\ie, deceive the detector to wrong predictions by strategically placing adversarial objects around the prohibited item.} However, simply extending existing physical attacks that work well on natural images to the context of X-ray images is non-trivial owing to the different imaging principles, \eg, the wavelengths of X-rays (\emph{0.001$\sim$10nm}) and visible light (\emph{390$\sim$780nm}) have a huge difference. More specifically, X-ray imaging is primarily conducted by utilizing material, thickness, and attenuation coefficients, meaning that the existing physical attacks designed for a visible light context (\eg, interference textures~\cite{wang2021dual} or patches~\cite{brown2017adversarial}) cannot be effectively applied to X-ray imaging. Thus, X-ray attacks should be considered a new type of attack problem in visually constrained scenarios with different wavelengths. In particular, we identify two key challenges impeding successful and feasible adversarial attacks in this scenario: (1) Color/texture fading. Due to its use of special imaging principles (\ie, beam intensity and attenuation rule), the X-ray scanning process eliminates most of the colors/textures and projects its outputs primarily based on item materials and shapes. Thus, the commonly used perturbations utilizing color disturbances will be removed by the X-ray scanning causing them to be ineffective. (2) Complex overlap. Luggage passed through an X-ray scanner often contains a large number of objects made of different materials, and overlap between these objects can degrade the attack performance; moreover, a successful attack should not rely on the occlusion of the prohibited item. Thus, when designing an adversarial object, it is necessary to consider the worst-case scenario (complex overlapping instances within the luggage), which increases the difficulty of the task.

To address the above problems, this paper proposes an adversarial attack approach called $\mathcal{X}$-\emph{Adv} to generate physically realizable adversarial attacks for X-ray prohibited item detection (as shown in Figure \ref{fig:frontpage}). As for the \emph{color/texture fading}, we generate physically realizable 3D objects with adversarial shapes, which enable our attacks to remain effective (since the shape cannot be altered after the X-ray imaging). To guide the design of the shape, we derive a differentiable converter that projects 3D objects into X-ray images so that we could update the shape of the object using the gradients of a surrogate white-box detector. As for the \emph{complex overlap}, we aim to find the locations that achieve strong attack ability even when occluded by other objects; moreover, we ensure that the placed adversarial objects do not overlap with the prohibited item. We thus introduce a policy-based strategy to search for the location that provides optimal attacking performance in the worst-case scenario. In summary, our $\mathcal{X}$-\emph{Adv} can generate adversarial objects by jointly optimizing the shapes and locations for X-ray attacks.

Extensive experiments in both the digital and physical world using multiple benchmarks against several detectors are conducted. Specifically, we first evaluate digital-world {attacks} on multiple benchmarks against both one-stage and two-stage detectors. We then successfully attack a commercial X-ray security inspection system in the real world by generating adversarial metal objects using a 3D printer. Finally, we present \final{the physical-world X-ray adversarial attack dataset XAD which contains 5,587 images (840 adversarial images)}. We hope this paper will draw more attention to the potential threats in safety-critical scenarios. Our \textbf{contributions} are:

\begin{itemize}
    \item To the best of our knowledge, this paper is the first work to study the feasibility of physical-world adversarial attacks in the visually-constrained X-ray imaging scenario.
    \item We propose the $\mathcal{X}$-\emph{Adv} to generate physically realizable adversarial metal objects for X-ray security inspection attacks by addressing the color fading and complex occlusion challenges.
    \item We conduct extensive experiments on several datasets in both digital- and physical-world settings, and the results demonstrate the effectiveness of our attack.
    \item We \final{present the physical-world X-ray adversarial attack dataset}, XAD, consisting of 5,587 images (840 adversarial images).
\end{itemize}

\section{Backgrounds and Related Work}

\quad\textbf{Prohibited Item Detection in X-ray Images.} X-ray imaging has been widely used due to its strong penetrative ability. In the X-ray security inspection scenario, inspectors usually adopt X-ray scanners to check passengers' luggage for the presence of prohibited items. A plethora of studies have been devoted to detecting prohibited items (\eg, pistols) in X-ray scanned luggage images using object detection methods \cite{wei2020occluded,Tao:ICCV21,wang2021towards,miao2019sixray} to detection performance.

In addition to the X-ray image detection methods, high-quality X-ray image datasets and benchmarks are also valuable for promoting the development of the research area. Though obtaining colorful X-ray images requires high computational costs, there are still some available open-source specialized datasets for X-ray security inspection. For instance, SIXray \cite{miao2019sixray} is a large-scale X-ray dataset containing millions of X-ray images collected from real-world subway stations. However, the images containing prohibited items are less than 1\%, and there is no bounding box annotation provided for object detection. Some high-quality X-ray datasets for object detection have also been made available. Wei \textit{et al.} \cite{wei2020occluded} first released the OPIXray dataset, which contains 8,885 artificially synthesized X-ray images of five categories of cutters. Tao \etal \cite{tao2021towards} proposed the HiXray dataset, comprising 45,364 images containing 102,928 prohibited items. All images from the dataset are collected from X-ray scanners in airports. Recently, Tao \etal \cite{Tao2022FSOD} further proposed the first few-shot object detection dataset in the X-ray security inspection scenario.


\textbf{Adversarial Attacks.} Adversarial examples are inputs with small perturbations, which are imperceptible to humans but can easily mislead DNNs into making incorrect predictions \cite{szegedy2013intriguing,goodfellow6572explaining}. Generally, we can classify them into digital and physical attacks. \emph{Digital attacks} usually generate adversarial perturbation at the pixel level across the whole input image. Szegedy \textit{et al.}\cite{szegedy2013intriguing} first defined adversarial examples and proposed L-BFGS attacks. By leveraging the gradient of the target model, Goodfellow \textit{et al.}\cite{goodfellow2014explaining} proposed FGSM to quickly generate adversarial examples. Since then, many types of adversarial attacks have been proposed, such as PGD \cite{madry2017towards}, DeepFool \cite{moosavi2016deepfool}, and JSMA \cite{papernot2016limitations}. However, due to their addition of global perturbations to the whole image, these attacks lack physical-world feasibility.

By contrast, \emph{physical attacks} aim to generate adversarial perturbations by perturbing the visual characteristics of real objects in the physical world. To achieve this goal, adversaries often generate adversarial perturbations in the digital world, then perform physical attacks by applying adversarial patches, painting adversarial camouflage, or directly creating adversarial objects in the real world \cite{brown2017adversarial,Eykholt_2018_CVPR,chen2018shapeshifter,wang2021dual}. Brown \etal \cite{brown2017adversarial} first proposed the adversarial patch by confining the perturbations into a local patch, which could then be printed to deceive the classification models. Eykholt \etal \cite{Eykholt_2018_CVPR} then modified the attacking loss function and generated strong adversarial attacks for real-world traffic sign recognition. Chen \etal \cite{chen2018shapeshifter} proposed Shapeshifter to attack a Faster R-CNN object detector in the physical world, specifically by attaching it to the STOP signs. In addition to the physical attacks on natural images (visible light domain), there also exist some preliminary studies on other \emph{visually constrained scenarios}. For example, Cao \etal \cite{cao2021invisible} investigated attacks in multi-sensor fusion scenarios, making adversarial examples invisible to both cameras and LiDAR. Recently, Zhu \etal \cite{zhu2021fooling, zhu2022infrared} attacked thermal infrared pedestrian detectors using small bulbs and special clothes. \revise{Mowery \etal \cite{mowery2014security} attacked a full-body X-ray scanner, while their proposed cyber-physical attacks did not aim at neural networks and are different from adversarial attacks.}

In summary, although numerous methods of physical attacks on natural images have been proposed, relatively little is known about the physical-world X-ray security inspection attack. This paper takes the first step to study physical-world adversarial attacks for X-ray security inspection.

\section{Threat Model}

\subsection{Problem Definition}
\label{sec:problem}
\textbf{Object detection.} An object detector $f_{\Theta}(\mathbf{I})\rightarrow \{\mathbf{b},\mathbf{c}\}^K$ with parameters ${\Theta}$, which takes an image $\mathbf{I} \in [0, 255]^n$ as input, outputs $K$ detection boxes with location $\mathbf{b}_k=[s_k, r_k, w_k, h_k]$ and confidence $c_k$. Moreover, $f$ applies a non-maximum suppression (NMS) operation to remove redundant bounding boxes. The formulation of the training is as follows:

\begin{equation}
    \min_{\Theta}\mathbb{E}_{(\mathbf{I}, \{\mathbf{y}_k, \mathbf{b}_k\})\sim \sD} \mathcal{L}(f_{\Theta}(\mathbf{I}), \{\mathbf{y}_k, \mathbf{b}_k\}),
\end{equation}

\noindent where $\mathcal{L(\cdot)}$ is the loss function that measures the difference between the output of the detector $f$ and the ground truth. $\mathbf{y}_k$ denotes the true label, and $\mathbf{b}_k$ denotes the true bounding box. In practice, the loss function is a weighted sum of the classification loss $\mathcal{L}_{cls}$ and location loss $\mathcal{L}_{loc}$:
\begin{equation}
    \min_{{\Theta}} \mathbb{E}_{(\mathbf{I}, \{\mathbf{y}_k, \mathbf{b}_k\})\sim \sD} [\mathcal{L}_{cls}(f^{cls}_{\Theta}(\mathbf{I}), \mathbf{y}_k) + \lambda \mathcal{L}_{loc}(f^{loc}_{\Theta}(\mathbf{I}), \mathbf{b}_k)].
\end{equation}

\textbf{Attacks on object detection.}  Given an object detector $f_{\Theta}$ and an input image $\mathbf{I} \in \sI$ with the ground truth label $\{\mathbf{y},\mathbf{b}_k\}$, an adversarial example $\mathbf{I}_{adv}$ satisfies the following:
\begin{equation}
f_{\Theta}(\mathbf I_{adv}) \neq \{\mathbf{y},\mathbf{b}_k\} \quad s.t. \quad \|\mathbf I-\mathbf I_{adv}\| \leq \epsilon,
\end{equation}
where $\|\cdot\|$ is a distance metric and commonly measured via $\ell_{p}$-norm ($p\in$\{1,2,$\infty$\}). Adversarial examples in visual recognition should also satisfy $\mathbf I_{adv}$ $\in$ $[0, 255]^n$. In this paper, we focus on deceiving the prediction class labels (\ie, $\rvy$).

\textbf{Physical attacks on X-ray prohibited item detection.} In this scenario, the items $\mathbf{X}=\{\mathbf{x}_1,...,\mathbf{x}_m\}$ in the luggage are scanned via an X-ray scanner to produce an X-ray image, where $\mathcal{R}$ denotes the process of generating a pseudo-color image depicted in Figure \ref{fig:frontpage} as $\mathbf{I}=\mathcal{R}(\mathbf{X})$). To perform physical attacks, we generate a 3D adversarial object $\mathbf{x}_{adv}$ with adversarial shapes $\mathbf{P}$ and place it at the proper location $\mathbf{C}$ in the luggage; the luggage is then scanned by the X-ray into image $\mathbf{I}_{adv}$, which could deceive the object detector $f_{{\Theta}}(\cdot)$, \ie, minimizing $\mathcal{M}$ that measures the performance of the detector:
\begin{equation}
    \min_{\mathbf{P}, \mathbf{C}} \mathcal{M}\left[f_{\Theta}(\mathcal{R}(\mathbf{x}_1,...,\mathbf{x}_m, \mathbf{x}_{adv}^{\mathbf{P}, \mathbf{C}}), \{\mathbf{y}_k, \mathbf{b}_k\})\right].
\end{equation}


\subsection{Challenges for X-ray Attacks}
Existing attacks mainly aim at the visible light domain by generating adversarial textures. However, it is highly challenging to directly apply these existing attacks to the X-ray domain. Specifically, we observe two main \textbf{challenges} as follows.

\textbf{{Challenge \ding{182}}}: \emph{The significant difference between imaging principles used in the visible light and X-ray contexts (\eg, different wavelengths).} We here first revisit the attenuation rule of X-ray photon beams. According to \cite{mccullough1974evaluation}, a narrow beam of X-ray photons with energy $E$ and initial photon intensity $I_0$, on passing through an absorber of small thickness $\Delta x$, will suffer a fractional decrease of intensity $\Delta I / I_0$ given by

\begin{equation}
\label{x-ray 1}
    \frac{\Delta I}{I_0}=-\mu (\rho, Z)\Delta x,
\end{equation}

\noindent where $\mu$ is the attenuation coefficient per unit length for an item made of a material of density $\rho$ and atomic composition $Z$. When the same photon beam passes through a certain absorber of finite thickness $x$, the intensity is given by

\begin{equation}
\label{x-ray 2}
    I = I_0\cdot exp(-\mu (\rho, Z)x).
\end{equation}

This attenuated intensity then will be received by sensors in X-ray scanners, according to which we can obtain the depth profile of the X-ray images. According to Equation \ref{x-ray 1} and \ref{x-ray 2}, we can conclude that an X-ray image is constructed primarily with reference to the material, the thickness of the object, and the properties of the light wave itself. Different from the perception of visible light images, X-rays tailor RGB space into a narrow color space, which means that common attacks that change pixel-wise textures will be ineffective for X-rays. To address this challenge, we need to optimize adversarial objectives to use non-color physical properties (\eg, shapes).

\textbf{{Challenge \ding{183}}}: \emph{Complex overlap due to the diversity of sampling scenarios and a massive number of luggage items in the X-ray security inspection context.} Placing the adversarial object directly on top of prohibited items would appear to be a simple attack method. However, this approach is infeasible in real-world applications, since luggage may be positioned randomly during X-ray scanning, and the overlap rate between adversarial objects and prohibited items under arbitrary sampling conditions is low. Moreover, this violates the definition of adversarial examples. To guarantee a feasible attack, the attacker should consider the worst-case scenario: that is, how to achieve an effective adversarial attack without occluding prohibited items, and with the overlapping of other objects.

\subsection{Adversarial Goals}
In this paper, we attempt to generate 3D adversarial objects with adversarial shapes to attack physical-world X-ray prohibited item detection models. As illustrated in Section \ref{sec:problem}, given an X-ray prohibited item detector $f_{\Theta}$ that takes an X-ray scanned image $\mathbf I$ as input, attackers aim to deceive $f_{\Theta}$ into making wrong predictions. \revise{This paper focuses on the more meaningful attack that deceives the detector to predict the wrong class labels rather than the wrong item locations. Specifically, we primarily study the untargeted attack, and the goal is to reduce the detection accuracy of detectors. Meanwhile, we also investigate the possibilities of the more difficult targeted attack, where we aim to force the detector predictions to the \texttt{Background} and make these prohibited items ``invisible'' (Section \ref{sec:discuss}). For the untargeted attack, the detector predicts any other labels that are different from the ground truth should be marked as a successful attack; while for the targeted attack, the prediction must match the assigned label.}

\subsection{Possible Attack Pathways}
Regarding adversarial attacks, one of the most important questions that should be answered is whether they are practical. For our $\mathcal{X}$-\emph{Adv} objects, they could be applicable to multiple X-ray image detection-related scenarios, \eg, security inspections in public hubs, and health examinations in hospitals. Using the $\mathcal{X}$-\emph{Adv} approach, adversaries could perform real-world attacks simply by generating an adversarial metal object by using 3D printers, then placing the item into their luggage or bags. \revise{The proposed attacks could make detectors yield wrong class predictions with low detection accuracy. Meanwhile, it is also possible for adversaries to conceal a prohibited item and make it ``invisible'' to the detectors, which can be achieved by simply modifying our attacking loss.}

\subsection{Adversary Constraints and Capabilities}
In considering the real-world X-ray security inspection scenario, we take comprehensive conditions into account and conduct both white-box and black-box attacks. In the white-box attack setting, the adversary has full access to the target model (\eg, architectures, weights), and is able to generate adversarial attacks directly based on its gradients. By contrast, the black-box attack setting is more practical; here, the adversary possesses only a little knowledge about the target model. For this setting, we assume that the target model and the source model are dealing with the same task and that the adversary performs transfer-based attacks. Specifically, the adversary first generates adversarial objects based on a white-box source model from a certain dataset; the adversary then prints the adversarial objects via a 3D printer in the real world; finally, adversaries could simply place adversarial objects in the luggage and attack the deployed X-ray security inspection model. Based on this, we could guarantee that all information of target models is unavailable to the attackers in black-box settings, which helps us to implement the strictest measures for simulating the physical scenarios. Moreover, to ensure our approach is more practical, the size of our adversarial objects should be small; thus the adversarial metal generated in this paper only takes up 1.78\% of the X-ray image.



\begin{figure*}[!t]
\vspace{-0.2in}
	\begin{center}
		\includegraphics[width=0.9\linewidth]{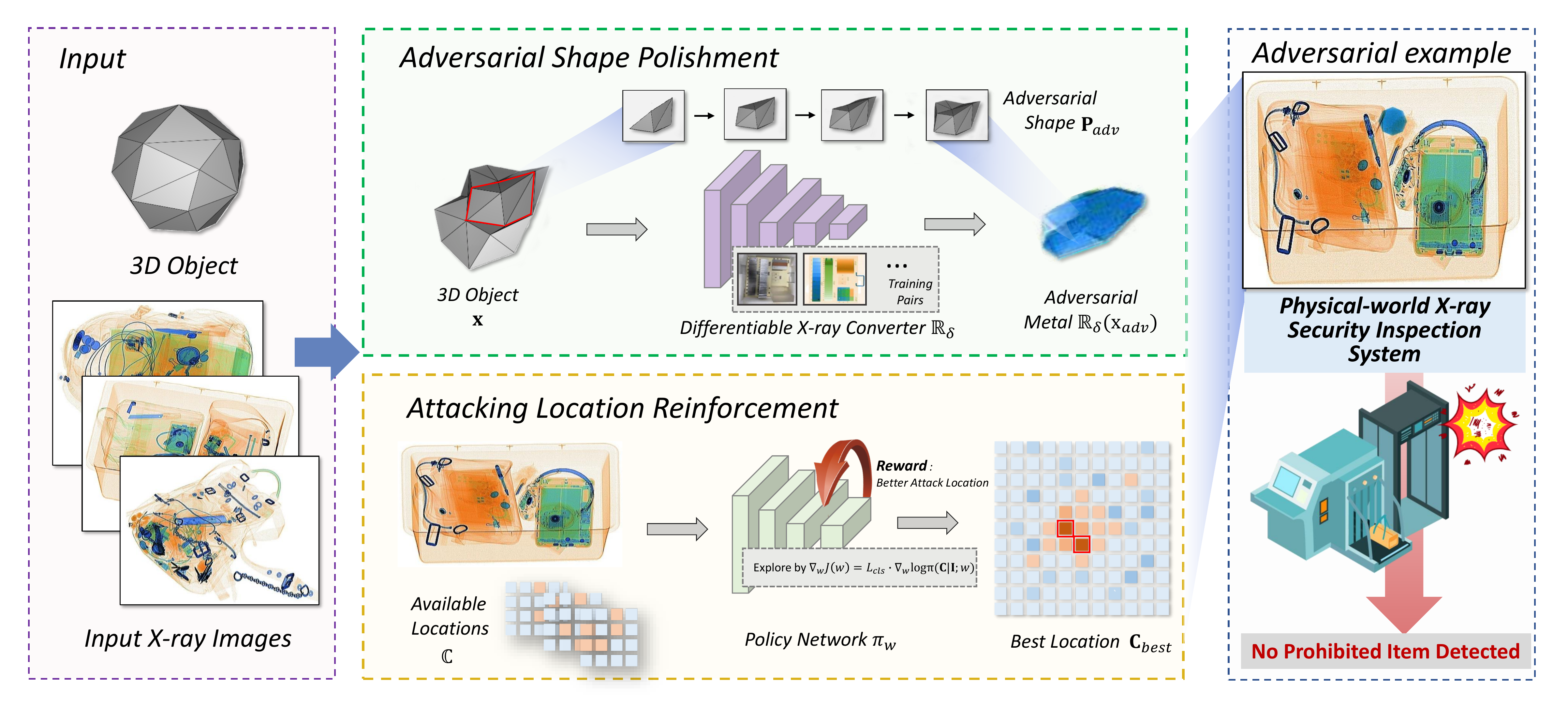}
	\end{center}
	\vspace{-0.15in}
	\caption{Illustration of \method approach. For the {color/texture fading} problem, we derive a differentiable converter that projects 3D objects into X-ray images; this allows us to generate 3D printable objects with adversarial shapes, which is X-ray projection invariant to X-ray imaging. We then introduce a policy-based algorithm to search for the optimal attacking location, which also shows high physical-world feasibility to address the complex occlusion problem. By jointly optimizing the combination of attack locations and shapes, our \method can generate physically realizable adversarial attacks for X-ray security inspection.}
	\label{fig:framework}
	\vspace{-0.1in}
\end{figure*}

\section{{$\mathcal{X}$}-\emph{Adv} Approach}
Selective Search~\cite{uijlings2013selective} proposes a heuristic strategy to discover potential objects from four perspectives: color similarity, texture similarity, shape similarity, and overlap similarity. Since it is based on the methods by which humans judge objects, this object discovery mechanism has been adopted in current deep-learning-based object detectors. Thus, to deceive the object detector, we can also optimize adversarial objectives from four perspectives: color, texture, shape, and overlap.

Due to the special nature of X-ray imaging principles and the diverse luggage sampling, physical-world adversarial attacks on X-ray security inspection should take the adversarial object's color/texture fading and complex overlapping problems into consideration. Therefore, this paper proposes $\mathcal{X}$-\emph{Adv} to generate physically realizable adversarial objects for X-ray security inspection (as illustrated in Figure \ref{fig:framework}). This approach simultaneously polishes adversarial shapes and reinforces attacking locations in the worst-case scenario (no overlapping) because the color space (\eg, color, texture) is not available.



\subsection{Adversarial Shape Polishment}
X-ray images, which are generated by X-ray security inspection machines from natural images, emphasize the shape and the material while neglecting the original color/textures of items. Thus, to successfully attack X-ray detectors, we generate objects with \emph{adversarial shapes} rather than \emph{adversarial textures} (since the adversarial colors/textures would be simply eliminated by the X-ray imaging pipeline, making such attacks ineffective). Accordingly, given a 3D object $\mathbf{x}$, we refine its visual characteristics $\mathbf{P}$ (\ie, shape) into adversarial shape $\mathbf{P}_{adv}$, such that the generated adversarial 3D object $\mathbf{x}_{adv}$ can attack the detector $f_{\Theta}(\mathcal{R}(\mathbf{x}_{adv}))$ after X-ray projection.

However, the X-ray imaging pipeline is highly complex and confidential and also varies significantly across different types of X-ray machines. Meanwhile, it is rather difficult to directly perform black-box query attacks since inspectors will not allow adversaries to query the system several times. Therefore, we propose a possible attack pathway in which we derive a differentiable X-ray converter to simulate the X-ray projection pipeline from 3D objects to 2D X-ray images and then perform gradient-based transfer attacks.

\final{However, the transformation from the depth $d$ of a scanned object to color images $g$ remains unknown. Based on the knowledge of X-ray machine vendors, the transformation process ($d\rightarrow I \rightarrow g$) can be simply represented by exponential functions, where the attenuated intensity $I$ of X-ray beams has an exponential relationship with the object depth $d$ (\cf Eqn. \ref{x-ray 2}) and the intensity $I$ to the color $g$ of X-ray images can be converted using a linear transformation. Therefore, we use the following exponential function to formulate the process:}

\begin{equation}
    g_m(d) = a\cdot exp(-b \cdot d) + q,
    \label{eqn:gmd}
\end{equation}

\noindent where $d$ indicates object depth, $m$ is the material, $g_m(d)$ represents the pixel value of color in a certain depth and material, and $a$, $b$, and $q$ are undetermined coefficients correlated to $m$, which will be calculated from real image sampling and regression fitting. \revise{We did not use DNNs for the transformation since it is too costly or even infeasible to collect sufficient data (\ie, different materials with diverse thicknesses) for DNN training (\emph{c.f.} Section \ref{sec:setup}). We use \texttt{HSV} color space rather than \texttt{RGB} because we found that regression in \texttt{HSV} space performs better in reducing the regression error (see Appendix \ref{sec:converter}).}

However, a depth image cannot represent a unique 3D object. Therefore, we use meshes as the format of our 3D adversarial object, given that meshes have been extensively used to parameterize 3D objects. Given an original mesh $\mathbf{x}_{ori}$, the coordinates on the \texttt{XY}-plane represent the shape of the image projected onto the 2D domain, while the coordinates in the \texttt{Z}-axis denote the depth (pixel value) of the image. Thus, we can optimize the shape of the 3D object by manipulating the coordinates in the mesh, then project the 3D mesh to a 2D depth image, and finally convert it to an X-ray image (the whole process is differentiable). The adversarial attack loss can be formalized by maximizing the classification loss $\mathcal{L}_{cls}$ of the target model, as follows:


\begin{equation}
    \mathcal L_{adv}(\mathbf X, \mathbf{x}_{adv}; f_{\Theta}, \mathbb{R_{\delta}})=\arg \max_{\mathbf{P}} \mathcal{L}_{cls}(f_\Theta (\mathbb{R}_{\delta}(\mathbf{X}, \mathbf{x}_{adv}^{\mathbf{P}})), \{\mathbf{y}_k, \mathbf{b}_k\}),
\end{equation}


\noindent where $\mathbb{R_{\delta}}$ denotes the differentiable converter in \revise{Eqn. \ref{eqn:gmd}} that can simulate the black-box X-ray scanning process $\mathcal{R}$. 

In practice, we overlap the original X-ray image $\mathbf{I}$ and the converted output $g(\cdot)$ by $\mathbf{I} \odot g_m(d(\mathbf{x}_{adv}^{\mathbf{P}}))$, where $d(\mathbf{x}_{adv}^{\mathbf{P}})$ refers to the \texttt{Z}-axis depth of adversarial mesh $\mathbf{x}_{adv}^{\mathbf{P}}$, and $\odot$ indicates pixel-wise multiplication.

\subsection{Attack Location Reinforcement}

In the X-ray security inspection scenario, there are often a vast number of items in the luggage. The simplest attack method is to attack the target detector by forcing a high overlap of adversarial objects and prohibited items. However, because the angle at which the bag will be scanned is uncontrollable in an X-ray detection scenario, the overlapping probability of adversarial targets and prohibited items are low, while adversarial objects are often occluded by other goods. This complex occlusion problem brings challenges to adversarial attacks. Therefore, we need to study effective attack algorithms in the \emph{worst-case} scenario, whereby the adversarial object does not cover prohibited items and is heavily occluded by other objects. Moreover, an appropriate location would increase the effectiveness of attacks, since our \method can only modify shapes while DNNs are more sensitive to texture \cite{hosseini2018assessing, shi2020informative, co2021universal}. 

Thus, to perform attacks under such a constrained scenario, we make full use of the location of the adversarial object and further improve the efficiency of our attacks by searching for the optimal attack locations. Accordingly, to achieve strong attacks, it is necessary to jointly consider the combination of attacking location and shape $(\mathbf{C}_{best}, \mathbf{P}_{best})$:

\begin{equation}
    \mathcal L_{adv}=\arg \max_{\mathbf{C}, \mathbf{P}} \mathcal{L}_{cls}(f_\Theta (\mathbb{R}_{\delta}(\mathbf{X}, \mathbf{x}_{adv}^{\mathbf{P}, \mathbf{C}})), \{\mathbf{y}_k, \mathbf{b}_k\}).
\end{equation}

Meanwhile, these two variables are mutually interactive, and the shape $\mathbf{P}$ is often influenced by the attack locations $\mathbf{C}$. If we determine the attack location $\mathbf{C}_{best}$, the adversarial shape $\mathbf{P}$ can be optimized by the gradient descent algorithm introduced in the previous subsection using the differentiable converter. However, there is no gradient information available for the attack location, which prevents us from optimizing the coordinates of adversarial objects. Also, calculating all possible conditions would result in unacceptably high computational costs. To tackle this problem, we apply a policy-based algorithm to search for the optimal attack location.

Inspired by \cite{zoph2016neural}, we use the REINFORCE algorithm \cite{williams1992simple} to introduce gradients between attack locations and the cost function. We define $\mathbb{C}$ as a finite area surrounding the prohibited item, or the ``available attacking area'', based on the ground truth bounding box, where we can place our adversarial objects. \revise{We simulate the common suitcases scanning scenario in the fixed top-down orientation and divide $\mathbb{C}$ into $N$ evenly spaced grids in the 2D space. In this scenario, the searching problem is relatively simple and the trajectory (state$\rightarrow$input, action$\rightarrow$location, reward$\rightarrow$loss) has only one timestep, and we use $N$ discrete actions to substitute location-choosing operations in a continuous area.} We define the policy network $\pi_{\mathbf{w}}$ with parameters $\mathbf{w}$, which receives the original image $\mathbf{I}$ as input and outputs the attacking location $\mathbf{C}$. The gradient of the objective function $J(\mathbf{w})$ with respect to $\mathbf{w}$ is shown as:

\begin{equation}
\label{eqn:reinforce}
    \nabla_{\mathbf{w}} J(\mathbf{w}) = G \cdot \nabla_{\mathbf{w}}\log \pi(\mathbf{C}|\mathbf{I}; \mathbf{w}),
\end{equation}
where $G$ refers to the reward of the policy. To enhance the feasibility in the physical world, we expect the locations of adversarial objects to vary, which can be quantified by the standard variance $\sigma_\mathbf{C}$. Therefore, $G$ consists of two components, attack capability, and location diversity, which can be written as:

\begin{equation}
\label{eqn:reward}
    G = \mathcal{L}_{cls}(f_\Theta (\mathbb{R}_{\delta}(\mathbf{X}, \mathbf{x}_{adv}^{\mathbf{P}_{ori}, \mathbf{C}})), \{\mathbf{y}_k, \mathbf{b}_k\}) + \alpha \cdot \sigma_\mathbf{C},
\end{equation}
where $\mathbf{P}_{ori}$ is the initial shape of the adversarial object, and $\alpha$ balances the two terms. With our policy-based searching algorithm, we can jointly optimize the location and shape of our adversarial objects, enabling us to perform efficient and effective physical-world attacks.

\subsection{Overall Optimization}

Based on the above discussions, the overall optimization function of our attacks $\mathcal{L}$ consists of the attack loss $\mathcal{L}_{adv}$ and perceptual loss $\mathcal{L}_{per}$, which can be written as follows:

\begin{equation}
\label{eqn:totalloss}
    \mathcal{L}(\mathbf{X},\mathbf{x}_{adv};f_{\Theta},\mathbb{R}_{\delta}) = \mathcal L_{adv}(\mathbf X, \mathbf{x}_{adv}; f_{\Theta}, \mathbb{R_{\delta}}) + \beta \mathcal{L}_{per}(\mathbf{x}_{adv}, \mathbf{x}_{ori}),
\end{equation}

\noindent where we append the adversarial attack loss with a perceptual loss $\mathcal{L}_{per}$ to ensure the physical feasibility of adversarial meshes, while $\beta$ is a coefficient to balance the two loss functions. Inspired by \cite{xiao2019meshadv, cao2019adversarial}, we further introduce a total variation loss into our perceptual loss to restrict the shape change as

\begin{equation}
\label{eqn:perc}
     \mathcal L_{perc}(\mathbf{x}_{adv}, \mathbf{x}_{ori}) = \frac{1}{|\mathbf{x}|} \sum_{V \in \mathbf{x}_{adv}} \sum_{v_i \in V} \sum_{v_q \in N(v_i)} ||\Delta v_i - \Delta v_q||_2^2,
\end{equation}

\noindent where $V$ is the vertex set of 3D adversarial meshes, $\Delta v_i$ indicates the perturbation distance of a certain vertex $v_i$ between $\mathbf{x}_{adv}$ and the original object $\mathbf{x}_{ori}$, and $N(v_i)$ refers to the vertices adjacent to $v_i$. The perceptual loss expects that a vertex will have similar perturbations to its neighbors, which avoids severe distortion of adversarial meshes. The pseudo-algorithm code of our \method can be found in Appendix \ref{pseudocode}.



\section{Experiments}
\subsection{Experimental Setup}
\label{sec:setup}

\quad\textbf{Datasets.}
We choose the commonly-used OPIXray \cite{wei2020occluded} and HiXray \cite{tao2021towards} datasets. Specifically, the OPIXray dataset has 7,109 images in the training set and 1,776 images in the test set, with five prohibited item categories (\eg, Folding Knife, Straight Knife). All data are scanned from an X-ray security inspection machine to reproduce the real-world scenario in public transportation hubs. The HiXray dataset has 36,295 images in the training set and 9,069 images in the test set, with eight prohibited item categories such as lithium battery, liquid, lighter, \etc{} \revise{Images from these datasets are captured and gathered from realistic sources, which contain diverse suitcases of different sizes and shapes (\eg, open trays, bags, luggage), and the prohibited items are surrounded randomly by other items of different materials (\eg, clothes, phones, laptops). Thus, experiments on them could better verify the effectiveness of our attack.}

\textbf{Target models.}
To verify the effectiveness of our attacks, we train both one-stage SSD \cite{Liu2016SSD} and two-stage Faster R-CNN \cite{girshick2015fast} to attack; we also attack the state-of-the-art and commonly used detectors in X-ray prohibited item detection scenario (DOAM \cite{wei2020occluded} and LIM \cite{tao2021towards}), where \revise{we achieve similar results on clean images compared to their original papers.} 

\textbf{Compared baselines.}
As discussed above, we are the first to study adversarial attacks for X-ray prohibited item detection, especially in the physical world. However, to better illustrate the superiority of our attacks, we transfer some adversarial attacks from prior works into the X-ray image scenario and compare our \method with them. Specifically, we use the original adversarial patch\cite{brown2017adversarial} (denoted as "AdvPatch") combined with our differentiable converter to generate 2D patches, which have no physical feasibility. As for 3D meshes, we apply meshAdv\cite{xiao2019meshadv} with a certain color of the adversarial patch (denoted as "meshAdv") as a baseline. We also apply vanilla adversarial objects without shape polishment and location reinforcement (denoted as "Vanilla") to examine the capability of the attacks above. Considering the cross-task domain gap, it is reasonable to expect that these comparison methods will not perform as well as their source works on the task at hand.

\textbf{Evaluation metrics.}
\revise{We select the most widely used metric, \ie, mAP, as the main evaluation metric. The mAP value depicts the overall performance according to precision and recall values, \ie, the area integral to the prediction precision ($\frac{TP}{TP+FP}$) and the prediction recall ($\frac{TP}{TP+FN}$) of object detection. Note that we set the IoU value (the intersection rate of the predicted border and the real border) as 50\%. In particular, the lower mAP values indicate better attack performance. For the untargeted attack, we use mAP to evaluate the attacking performance; for the targeted attack, besides mAP, we also report the False Negative (FN) values with confidence as 0.8 (the higher the better).}

\textbf{Implementation details.}
\revise{We define the size of the adversarial object as 20$\times$20 square pixel and the number of objects as 4, which takes around 2\% of the whole image.} \textit{More details are shown in Appendix \ref{sec:moresetting}.} All the codes are implemented with PyTorch. For all experiments, we conduct the training and testing on an NVIDIA GeForce RTX 2080Ti GPU cluster. 


\textbf{X-ray converter.} We obtain the coefficient of the X-ray converter using a commercial AT6550 X-ray scanner. In practice, we have scanned 8 thicknesses (\emph{0.2$\sim$8mm}) of iron objects, 22 thicknesses (\emph{1$\sim$60mm}) of aluminum objects, and 6 thicknesses (\emph{60$\sim$120mm}) of plastic objects using our X-ray machine. Then we sampled their color under X-ray images. We use Eqn. \ref{eqn:gmd} as the convert function, the coefficients of which are acquired from regression fitting.

\subsection{Digital-world Attacks}

In this part, we evaluate our \method in the digital world under both white-box and black-box settings. Specifically, for the white-box setting, we generate the adversarial object based on the model, then test its attacking ability on the same model; for the black-box setting, we first optimize the adversarial object on one model, then test its attack performance on other models via transfer-based attack. In more detail, we employ 4 models in the digital-world experiments including both one-stage and two-stage detectors, and the white-box attack results on OPIXray are shown in Table \ref{table:digital}. \emph{More results on HiXray and black-box attacks can be found in Appendix \ref{result}.} From the results, we can \textbf{identify}:

\ding{182} Despite having eliminated most of the colors and textures in the X-ray images, the adversarial attacks still pose challenges in the X-ray prohibited item detection scenario. For example, on the OPIXray dataset against DOAM, the clean mAP is 74.02\%, while the mAP value drops significantly to \textbf{23.05\%} after being attacked by our $\mathcal{X}$-\emph{Adv}. It should be noted that this observation can be made for all employed models: the observed average mAP degeneration is about \textbf{50\%} on OPIXray and \textbf{30\%} on HiXray. Moreover, our \method outperforms other baselines by large margins.


\ding{183} It should be noted that \method seems to fail in some categories of HiXray, \eg, laptops. We hypothesize that the reason lies in the characteristic of the target object. Laptops usually occupy large proportions of an image, while our patch is much smaller than these objects. Therefore, detector models can obtain much more information about objects like laptops, which supports correct classification.

\ding{184} Moreover, it is important to note that the vanilla patch could not successfully attack the detector, which indicates that the observed vulnerabilities of these models are not the result of poorly trained detectors.

\begin{table}[!t]
\begin{center}
\vspace{-0.15in}
\caption{Digital-world white-box attacks on OPIXray. ``FO'', ``ST'', ``SC'', ``UT'', and ``MU'' represent Folding Knife, Straight Knife, Scissor, Utility Knife, and Multi-tool Knife.}
\label{table:digital}
\vspace{-0.1in}

\subtable[SSD]{
\footnotesize
\setlength{\tabcolsep}{2.8mm}{
\begin{tabular}{@{}ccccccc@{}}
\toprule
\multirow{2}{*}{Setting} & \multirow{2}{*}{mAP} & \multicolumn{5}{c}{Categories}                                                   \\ \cmidrule(l){3-7} 
                         &                      & FO             & ST            & SC             & UT             & MU            \\ \midrule
Clean    & 72.23          & 78.37          & 37.82         & 92.49          & 69.58          & 82.87          \\
Vanilla  & 61.46          & 71.51          & 17.86         & 90.20          & 52.45          & 75.29          \\
MeshAdv  & 52.77          & 61.82          & 10.20         & 83.72          & 40.54          & 67.59          \\
AdvPatch & 40.91          & 47.19          & 5.86          & 74.83          & 25.48          & 51.21          \\
\method  & \textbf{19.20} & \textbf{24.11} & \textbf{1.46} & \textbf{44.48} & \textbf{12.59} & \textbf{13.37} \\ \bottomrule
\end{tabular}
}
}

\subtable[Faster R-CNN]{
\footnotesize
\setlength{\tabcolsep}{2.8mm}{
\begin{tabular}{@{}ccccccc@{}}
\toprule
\multirow{2}{*}{Setting} & \multirow{2}{*}{mAP} & \multicolumn{5}{c}{Categories}                                                   \\ \cmidrule(l){3-7} 
                         &                      & FO             & ST            & SC             & UT             & MU            \\ \midrule
Clean    & 64.92          & 60.90          & 37.19         & 89.74          & 66.82          & 69.96         \\
Vanilla  & 53.05          & 53.13          & 20.75         & 85.69          & 49.76          & 55.93         \\
MeshAdv  & 49.49          & 44.26          & 17.48         & 81.70          & 44.03          & 59.99         \\
AdvPatch & 50.19          & 52.67          & 15.88         & 84.03          & 42.26          & 56.13         \\
\method  & \textbf{23.33} & \textbf{26.62} & \textbf{3.44} & \textbf{62.91} & \textbf{15.33} & \textbf{8.36} \\ \bottomrule
\end{tabular}
}
}

\subtable[DOAM]{
\footnotesize
\setlength{\tabcolsep}{2.8mm}{
\begin{tabular}{@{}ccccccc@{}}
\toprule
\multirow{2}{*}{Setting} & \multirow{2}{*}{mAP} & \multicolumn{5}{c}{Categories}                                                   \\ \cmidrule(l){3-7} 
                         &                      & FO             & ST            & SC             & UT             & MU            \\ \midrule
Clean                    & 74.02                & 78.92          & 40.88         & 95.65          & 74.08          & 80.55         \\
Vanilla                  & 67.79                & 74.26          & 32.57         & 91.37          & 63.41          & 77.34         \\
MeshAdv                  & 56.36                & 60.09          & 23.04         & 86.87          & 47.11          & 64.68         \\
AdvPatch                 & 42.04                & 45.57          & 9.41          & 81.19          & 26.44          & 47.60         \\
\method                  & \textbf{23.05}       & \textbf{18.40} & \textbf{4.05} & \textbf{64.80} & \textbf{18.57} & \textbf{9.45} \\ \bottomrule
\end{tabular}
}
}

\subtable[LIM]{
\footnotesize
\setlength{\tabcolsep}{2.8mm}{
\begin{tabular}{@{}ccccccc@{}}
\toprule
\multirow{2}{*}{Setting} & \multirow{2}{*}{mAP} & \multicolumn{5}{c}{Categories}                                                   \\ \cmidrule(l){3-7} 
                         &                      & FO             & ST            & SC             & UT             & MU            \\ \midrule
Clean                    & 73.07                & 79.01          & 36.04         & 94.73          & 72.94          & 82.62         \\
Vanilla                  & 66.44                & 73.58          & 22.78         & 93.08          & 65.17          & 77.62         \\
MeshAdv                  & 59.60                & 65.56          & 19.70         & 87.27          & 52.26          & 73.20         \\
AdvPatch                 & 49.69                & 54.16          & 14.66         & 80.35          & 35.72          & 63.55         \\
\method                  & \textbf{22.46}       & \textbf{31.64} & \textbf{4.28} & \textbf{52.59} & \textbf{16.65} & \textbf{7.13} \\ \bottomrule
\end{tabular}
}
}

\end{center}

\vspace{-0.4in}

\end{table}

The results for the black-box setting in Appendix \ref{result} show consistent phenomena. In summary, the digital-world evaluations demonstrate that our \method could successfully attack the X-ray prohibited item detectors and outperform other baselines by large margins.


\subsection{Physical-world Attacks}
\label{sec:phy}
In this part, we further investigate the X-ray prohibited item detection model robustness in the physical-world setting.

\begin{figure}[!t]
	\begin{center}
		\includegraphics[width=1.0\linewidth]{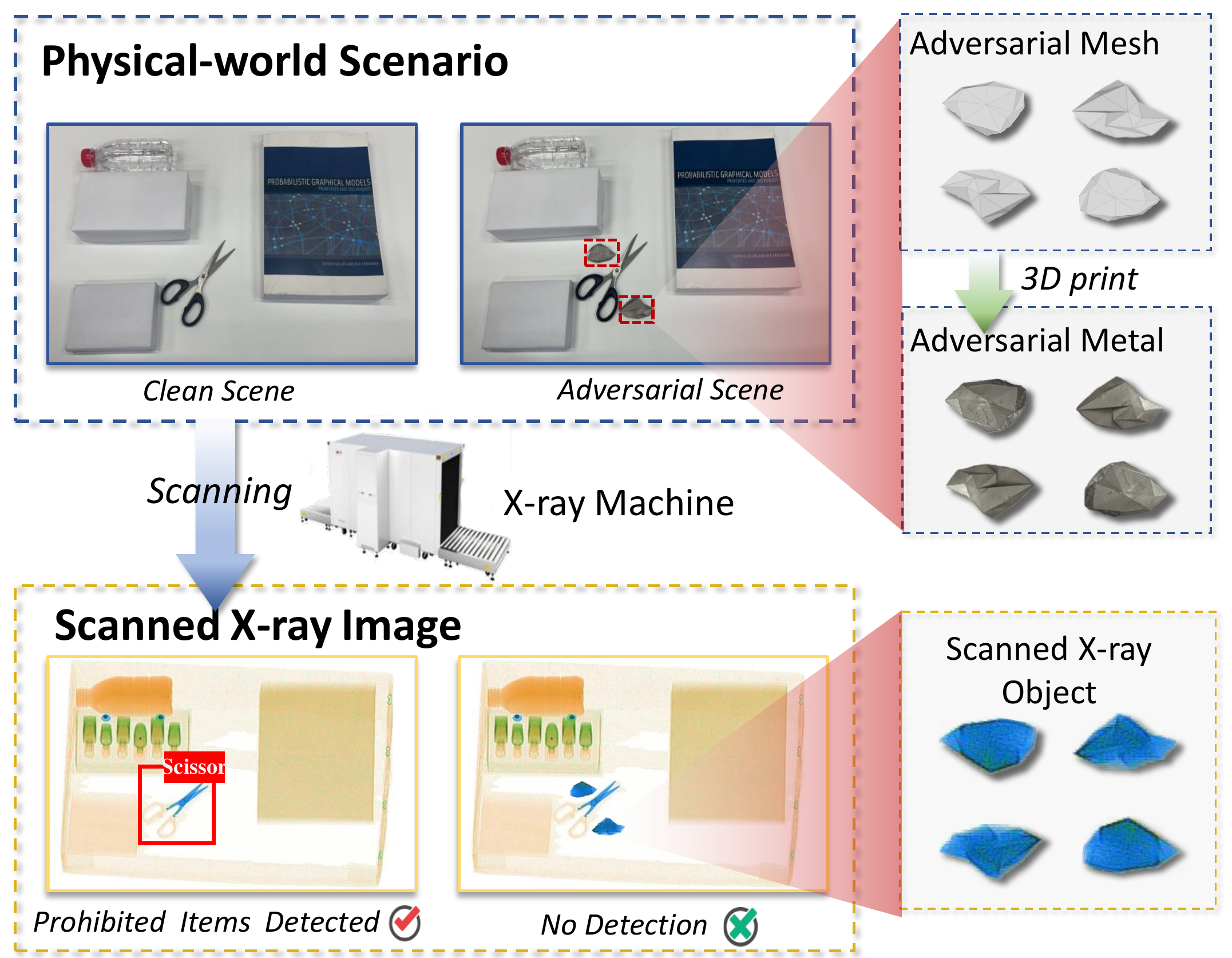}
	\end{center}
	\vspace{-0.15in}
	\caption{Illustration of the physical-world attacking pipeline. $\mathcal{X}$-adv first generates adversarial meshes (3D objects); we then print these meshes into metal objects using 3D printers; when scanned by X-ray scanners, these metal objects will become adversarial patches in the resulting X-ray images.}
    \label{fig:phy-pipeline}
	\vspace{-0.2in}
\end{figure}

We first illustrate the \textbf{attack pipeline} for our physical-world attacks (see Fig \ref{fig:phy-pipeline}). In detail: \ding{182} we first generate adversarial objects using our \method based on a white-box pre-trained DOAM target model; \ding{183} we then transform the adversarial objects from 3D mesh format into STL format so that we can use a third-party 3D printer to print these 3D objects in metal; \ding{184} we then put our adversarial objects into the fabric/plastic box with other items and employ several workers to scan them into X-ray images using a commercial AT150180B X-ray scanner (which is commonly used in the train station and airport security checkpoints); \ding{185} finally, we test our physical-world adversarial examples (X-ray images) on black-box X-ray prohibited item detection models, specifically, DOAM, LIM, and Faster R-CNN, which are trained on the physical-world dataset proposed in Section \ref{sec:xad}. \revise{Note that, we use the commercial X-ray scanners but cannot use their detection backend because these models/strategies are business secrets, however, we adopt a similar black-box Faster R-CNN. During the experiments, we have no access to and prior knowledge of these target detectors and X-ray scanners.}

\begin{figure}[!t]
	\begin{center}
	\vspace{-0.15in}	\includegraphics[width=\linewidth]{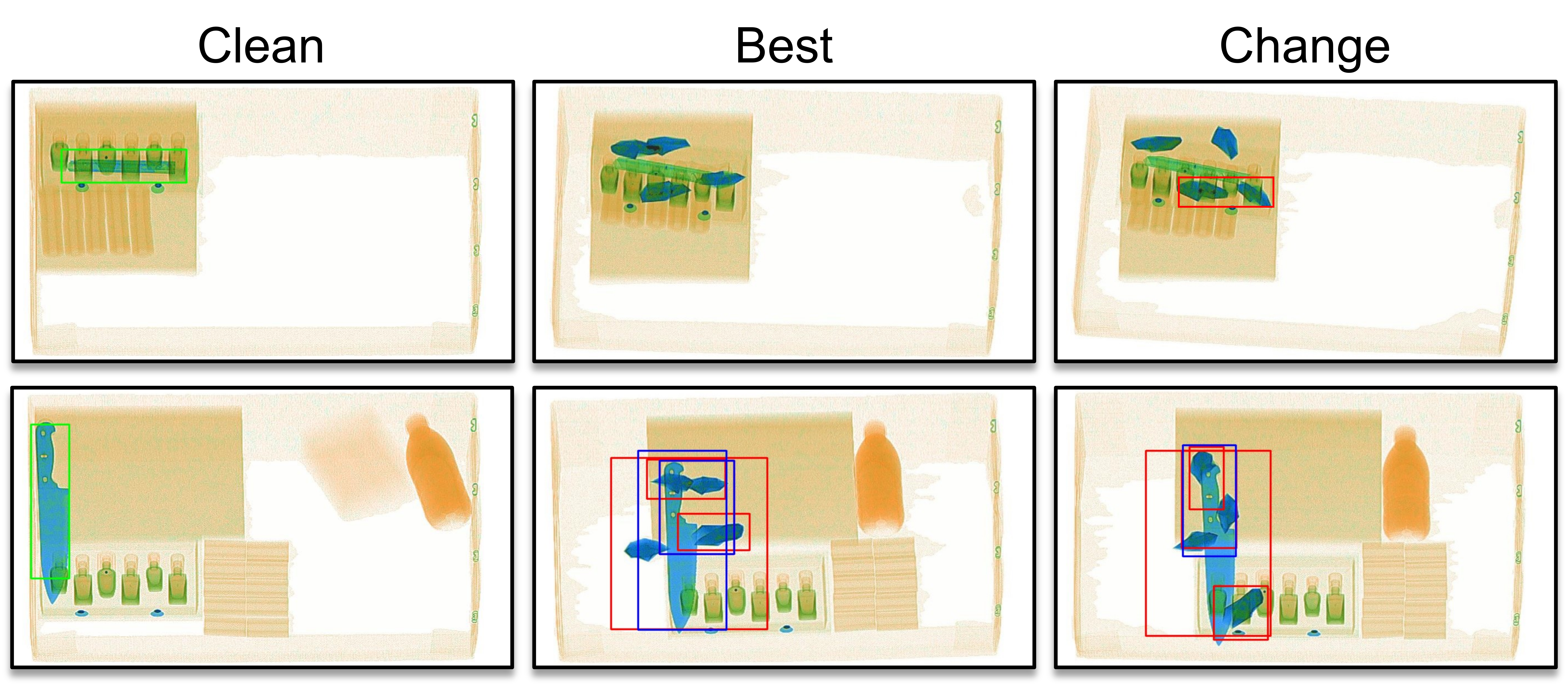}
	\end{center}
	\vspace{-0.2in}
	\caption{Detection results of some X-ray images in our physical-world experiments \revise{(we choose images with fewer items for better visualization)}. \textcolor{green}{Green boxes} indicate correct classes and suitable locations; \textcolor{blue}{blue boxes} represent correct classes in incorrect locations; \textcolor{red}{red boxes} indicate incorrect classes. We only show detection boxes with confidence >10\%.}
	\label{fig:physical2}
	\vspace{-0.2in}
\end{figure}

Specifically, we use \method to generate 16 adversarial metal objects and then print them in \texttt{iron} using a 3D printer. We collect items (\eg, laptops, headphones, bags) from our staff and students under their grants. In total, we collected 80 adversarial X-ray images as the test set; some physical-world clean and adversarial X-ray images can be found in Figure \ref{fig:physical2}. Note that all X-ray images are collected without personal information to avoid privacy leakage. To assess the real-world feasibility of our attacks, in addition to having our X-Adv search for the best possible attack location (denoted as ``Physical best''), \revise{we also impose two attack settings to better simulate the physical-world dynamic environment of items movement in luggage: (1) slight transformations and (2) random placement. Specifically, slight transformations add shift (random 10 pixels in each direction) and rotation (-30$^\circ$$\sim$+30$^\circ$) to adversarial objects (denoted as ``Physical change''), while random placement randomly places the adversarial objects in the entire suitcase (denoted as ``Physical random''). For better comparison, we also provide the results of the 80 images using digital-world attacks (denoted as ``Digital attack'') and the physical-world results on clean examples (denoted as ``Clean'').}

\begin{table}[!t]
\begin{center}
\caption{Physical-world attack experiments on different detection models. More results are shown in Appendix \ref{result}.}
\vspace{-0.15in}
\label{table:physical}
\subtable[DOAM]{
\footnotesize
\setlength{\tabcolsep}{3.0mm}{
\begin{tabular}{@{}cccccc@{}}
\toprule
\multirow{2}{*}{Setting} & \multirow{2}{*}{mAP} & \multicolumn{4}{c}{Categories} \\ \cmidrule(l){3-6} 
                         &                      & SC    & FO    & ST     & UT    \\ \midrule
Clean                    & 91.35                & 84.17 & 98.05 & 100.00 & 83.18 \\
Digital attack           & 30.28                & 67.54 & 2.15  & 50.73  & 0.69  \\
Physical best            & 33.16                & 66.33 & 18.35 & 44.48  & 3.46  \\
Physical change          & 50.97                & 74.13 & 42.19 & 55.92  & 31.63 \\
\revise{Physical random} & \revise{76.17}       & \revise{76.06} & \revise{79.19} & \revise{85.33}  & \revise{64.10} \\ \bottomrule
\end{tabular}
}
}

\subtable[Faster R-CNN]{
\footnotesize
\setlength{\tabcolsep}{3.0mm}{
\begin{tabular}{@{}cccccc@{}}
\toprule
\multirow{2}{*}{Setting} & \multirow{2}{*}{mAP} & \multicolumn{4}{c}{Categories} \\ \cmidrule(l){3-6} 
                         &                      & SC    & FO    & ST     & UT    \\ \midrule
Clean                    & 95.35                & 94.00 & 100.00 & 92.66 & 94.75 \\
Digital attack           & 27.18                & 44.77 & 0.31   & 50.63 & 13.00 \\
Physical best            & 24.67                & 62.88 & 2.26   & 23.03 & 10.53 \\
Physical change          & 57.38                & 85.84 & 35.45  & 72.16 & 36.07 \\
\revise{Physical random} & \revise{75.57}       & \revise{93.00} & \revise{56.03} & \revise{88.95}  & \revise{64.29}
\\ \bottomrule
\end{tabular}
}
}

\end{center}

\vspace{-0.25in}

\end{table}

From Table \ref{table:physical}, we can conclude that the physical attacking ability of the proposed \method has a significant impact on detection accuracy, \ie, the mAP value of DOAM on physical clean samples is 91.35\%, while the mAP values on the sampled adversarial samples are 33.16\% on ``Physical best'', 50.97\% on ``Physical change'', and \revise{76.17\% on ``Physical random''}, which are lower than the results on the clean counterparts. This observation also indicates that the safety problem of X-ray prohibited item detection is worth studying from a practical perspective. Moreover, we also observe that the attacking ability of ``Physical best'' is stronger than that of ``Physical change'' (lower mAP), which supports our motivation to search for the critical attack position. Furthermore, compared to the digital-world attack results, the physical-world attack results are weaker; we speculate this is because of the digital-physical domain gap \cite{eykholt2018robust, wang2021dual}.

\subsection{Ablation Studies}

In this section, we investigate the key factors that might impact the attack ability of our \method, thereby providing comprehensive insights and promoting a deeper understanding of our strategy. In brief, we conduct thorough ablations on several factors. All the experiments conducted in this part use the DOAM target model on OPIXray and HiXray datasets.

\revise{\textbf{Attack locations.} Here, we investigate three additional location-searching strategies on the attack performance, \ie, fixed position (denoted as ``Fix''), random positions (denoted as ``Random''), and greedy-search-based positions (denoted as ``Greedy''). Our proposed attacking location search strategy is denoted as ``Reinforce''. For Fix, we place the adversarial objects on the corners of the prohibited items; for Random, we place the adversarial objects randomly around the prohibited items; for Greedy, we first greedy-search the strongest attack locations that maximize $\mathcal{L}_{cls}$ by placing one original object at each location, then optimize the adversarial objects at the corresponding locations. The experimental results on OPIXray and HiXray can be found in Table \ref{table:place}, where we can observe that among all 4 attacking location searching strategies, the result under our ``Reinforce'' setting shows the strongest adversarial attacking performance.} 

\revise{Moreover, we study a more limited setting where the adversary could stick an adversarial object on the prohibited item. In particular, we add experiments on OPIXray against DOAM, where we put a 32$\times$32 iron rectangle or a 40$\times$40 adversarial object generated by \method on top of the target object. The attack performance (49.48 mAP and 25.28 mAP) is still worse than our original position-searching strategy (23.05 mAP). Note that directly placing the target object into an iron box or hiding it with iron plates could make it disappear, which can be easily identified in practice. Meanwhile, this would also violate the definition of adversarial attacks (cover the salient parts and change its semantics). The goal of this experiment is to illustrate the importance of location-searching.}

\revise{The above studies demonstrate that avoiding overlap and occlusion can increase the attack capability.}

\begin{figure}[!t]
\vspace{-0.1in}
    \centering
    \subfigure[OPIXray]{
        \includegraphics[width=0.47\linewidth]{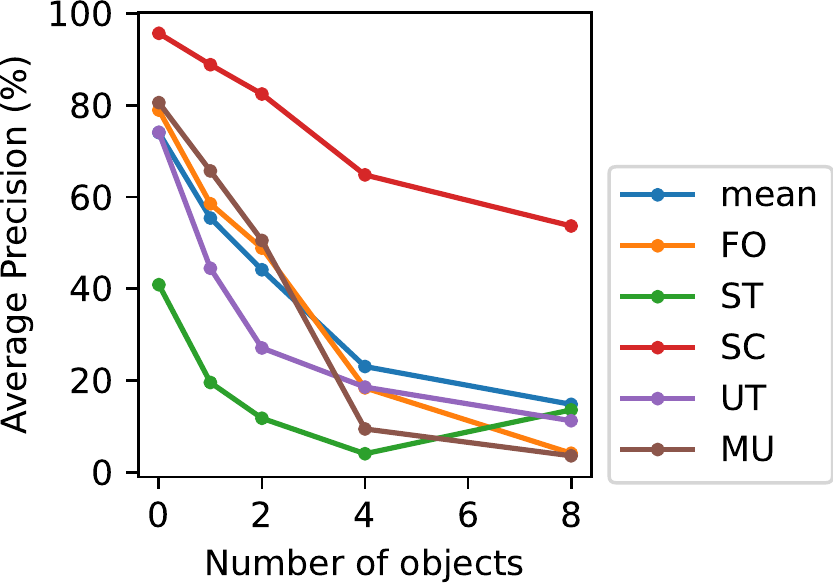}
    }
    \subfigure[HiXray]{
        \includegraphics[width=0.47\linewidth]{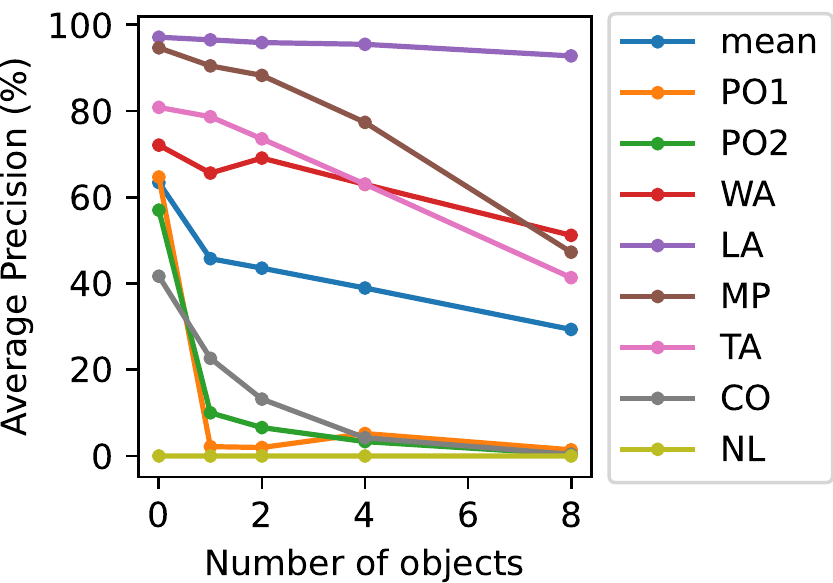}
    }
    
    \vspace{-0.15in}
    \caption{Ablations on the numbers of adversarial objects.}
    \vspace{-0.1in}
    \label{fig:number}
\end{figure}

\textbf{Number of objects.} Regarding the number of adversarial objects, we study whether the attack ability of the adversarial objects differs when this number is changed. Thus, we set the number of the adversarial objects to 1, 2, 4, and 8 respectively, while keeping the total area of each setting the same. The results can be found in Figure \ref{fig:number}. As the results show, more objects usually result in better attack performance. However, too many adversarial objects will introduce an additional cost in terms of physical feasibility. In practice, we set the number of objects as 4.

\subsection{Discussions and Analysis}
\label{sec:discuss}

In this part, we provide more detailed discussions and analysis on the attack ability and physical feasibility of \method. All the experiments conducted in this part are using the DOAM target model based on OPIXray and HiXray datasets.

\begin{table}[!t]
\vspace{-0.15in}
\begin{center}
\caption{Ablation studies on different attack locations. Our strategy achieves the best attack performance.}
\label{table:place}
\subtable[OPIXray]{
\footnotesize
\setlength{\tabcolsep}{2.9mm}{
\begin{tabular}{@{}ccccccc@{}}
\toprule
\multirow{2}{*}{Setting} & \multirow{2}{*}{mAP} & \multicolumn{5}{c}{Categories}                                                   \\ \cmidrule(l){3-7} 
                         &                      & FO             & ST            & SC             & UT             & MU            \\ \midrule
Fix                      & 51.64                & 55.54          & 18.22         & 82.16          & 39.89          & 62.38         \\
Random                   & 38.11                & 40.54          & 8.39          & 76.77          & 26.82          & 38.01         \\
\revise{Greedy}          & \revise{29.38}       & \revise{28.02} & \revise{5.02} & \revise{65.46} & \revise{20.21} & \revise{28.19} \\ 
Reinforce                & \textbf{23.05}       & \textbf{18.40} & \textbf{4.05} & \textbf{64.80} & \textbf{18.57} & \textbf{9.45} \\ \bottomrule
\end{tabular}
}
}

\subtable[HiXray]{
\resizebox{\linewidth}{!}{
\begin{tabular}{@{}cccccccccc@{}}
\toprule
\multirow{2}{*}{Setting} & \multirow{2}{*}{mAP} & \multicolumn{8}{c}{Categories}                                                                                                    \\ \cmidrule(l){3-10} 
                         &                      & PO1           & PO2           & WA             & LA             & MP             & TA             & CO            & NL            \\ \midrule
Fix                      & 44.68                & 10.48         & 8.95          & 69.06          & 96.42          & 88.76          & 74.69          & 9.04          & \textbf{0.00} \\
Random                   & 41.98                & 8.41          & 6.37          & 66.05          & 95.74          & 82.74          & 68.63          & 7.93          & \textbf{0.00} \\
\revise{Greedy}          & \revise{40.19}  & \revise{5.77}  & \revise{4.14} & \revise{64.88} & \revise{95.47} & \revise{80.44} & \revise{65.76}  &  \revise{5.06} & \revise{\textbf{0.00}} \\
Reinforce                & \textbf{38.96}       & \textbf{5.21} & \textbf{3.33} & \textbf{63.00} & \textbf{95.49} & \textbf{77.38} & \textbf{63.05} & \textbf{4.22} & \textbf{0.00} \\ \bottomrule
\end{tabular}
}
}

\end{center}

\vspace{-0.1in}

\end{table}

\begin{figure}[!t]
    \centering
    \subfigure[Materials]{
        \label{fig:material}
        \includegraphics[width=0.47\linewidth]{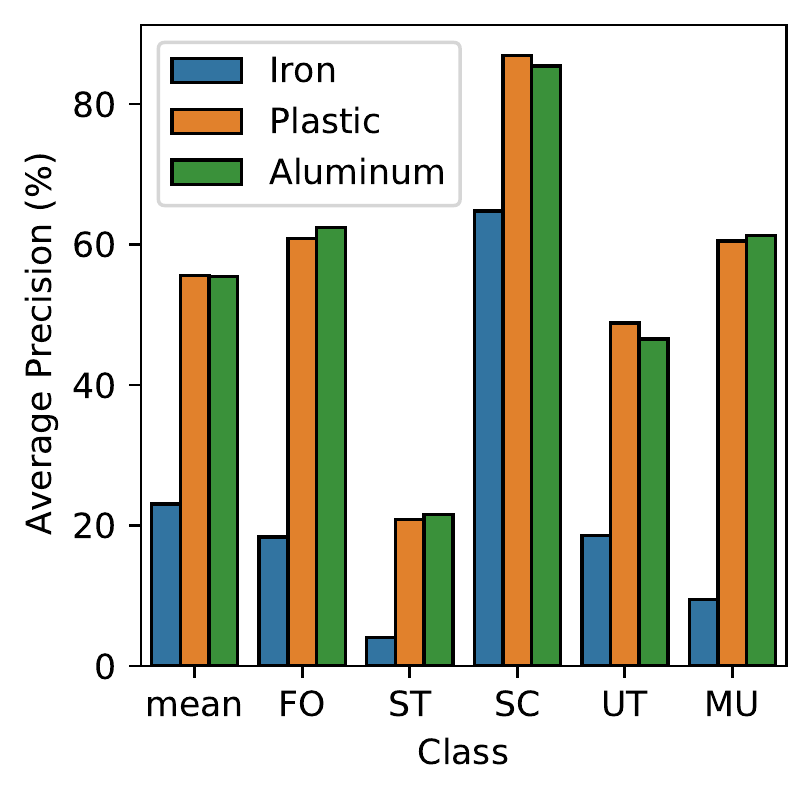}
    }
    \subfigure[Targeted/untargeted]{
        \label{fig:target}
        \includegraphics[width=0.47\linewidth]{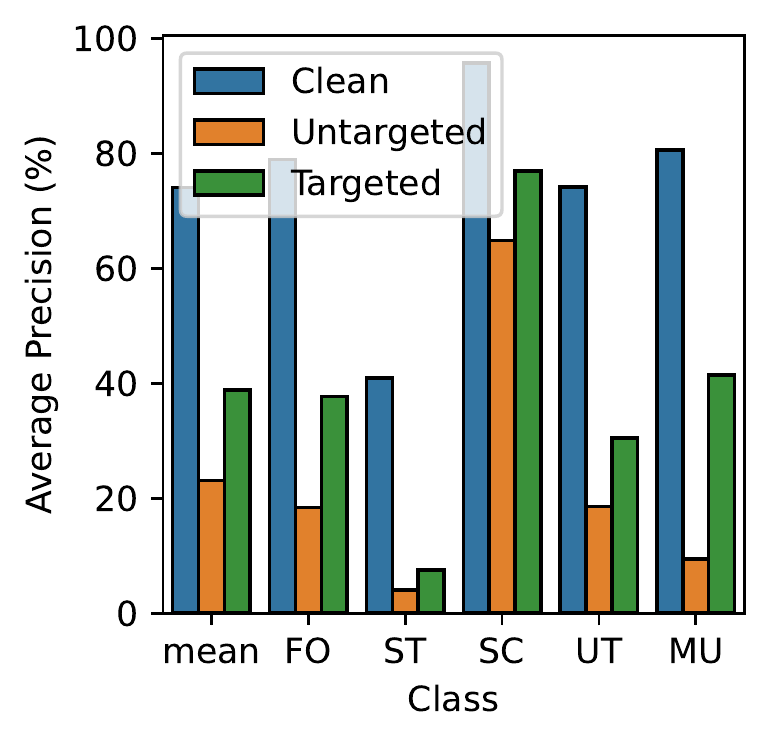}
    }
    \vspace{-0.15in}
    \caption{Results using DOAM on OPIXray: (a) different materials, (b) targeted and untargeted adversarial attacks.}
    \vspace{-0.1in}
\end{figure}

\textbf{Object materials.} Different materials are rendered on X-ray images in different colors; therefore, it is necessary to investigate their possible influence on the final attack ability of the generated adversarial objects. To this end, we select three kinds of colors (materials), namely blue, green, and orange, which respectively correspond to three materials that commonly appear in the luggage, \ie, iron, aluminum, and plastic. The results on OPIXray can be found in Figure \ref{fig:material}. It is clear that the generated adversarial objects with blue colors (iron) show stronger attack ability, \ie, lower mAP values. For instance, on the OPIXray dataset, the mAP value of the iron adversarial objects is \textbf{23.05\%}, while that of the green (aluminum) ones is 55.44\%, and that of the orange (plastic) ones is 55.61\%. We believe that this observation is reasonable since prohibited items (such as knives and guns) tend to be made of metal, thus adversarial objects made of similar materials (and rendered in similar colors) will have a higher attack ability. On the OPIXray dataset, the prohibited items are knives, meaning that blue (iron) outperforms other colors significantly. \emph{Results on HiXray can be found in Appendix \ref{result}.}

\begin{figure}[!t]
    \centering
    \includegraphics[width=1.0\linewidth]{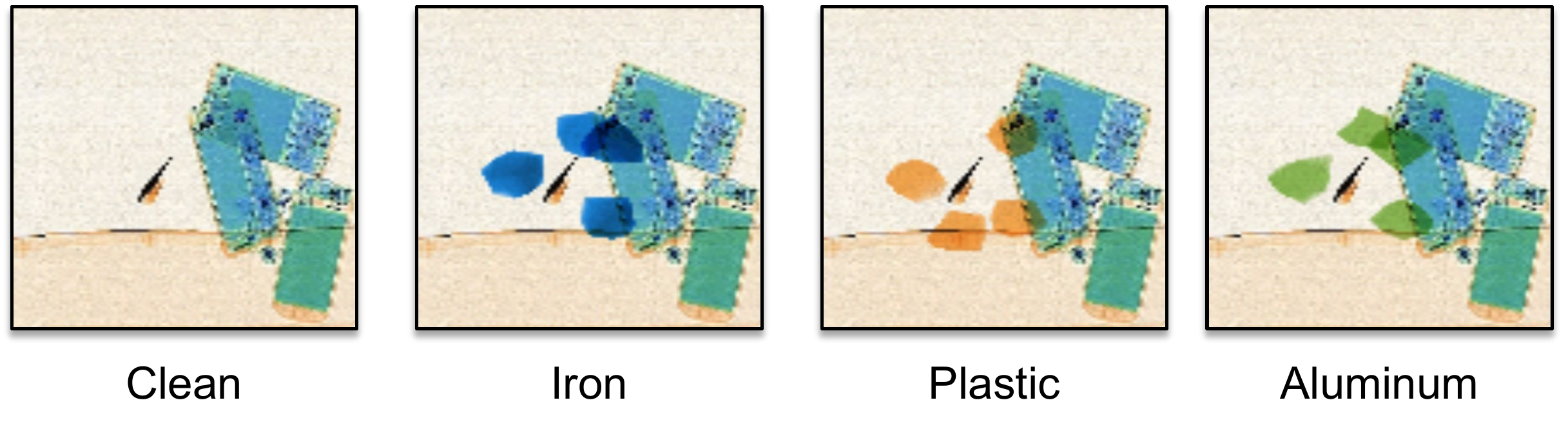}
    \vspace{-0.2in}
    \caption{Visualization of adversarial objects with different materials/colors (\ie, blue, orange, and green).}
    \vspace{-0.1in}
    \label{fig:color}
\end{figure}

\revise{\textbf{Targeted adversarial attacks.} As shown in Eqn. \ref{eqn:totalloss}, our \method maximizes the classification loss $\mathcal L_{cls}$ of the detector to output wrong classes, which is the \textit{untargeted attacks}. In addition to the untargeted attack, we here further design another reasonable attack strategy, \ie, perform attacks to evade the detector by confusing the predictor to classify the object of interest as \texttt{Background}. Thus, we apply \textit{targeted attacks} and set the target label of attacks to \texttt{Background} (one of the classes for detection). Specifically, we substitute $\mathcal L_{cls}$ with a cross-entropy loss between the confidence of all predicted boxes and the background class. Since the background is the 0-th class of object detectors, performing attacks that mislead all boxes into the background class can also reduce the number of predicted boxes.}

\begin{figure}[!t]
    \centering
    \subfigure[TP]{
        \includegraphics[width=0.47\linewidth]{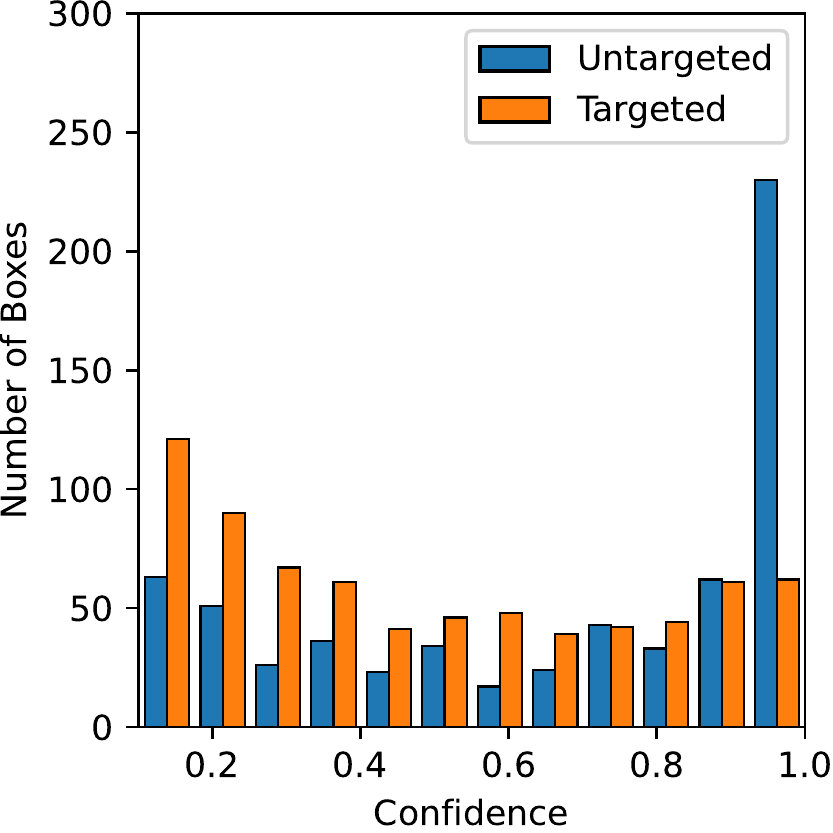}
    }
    \subfigure[FP]{
        \includegraphics[width=0.47\linewidth]{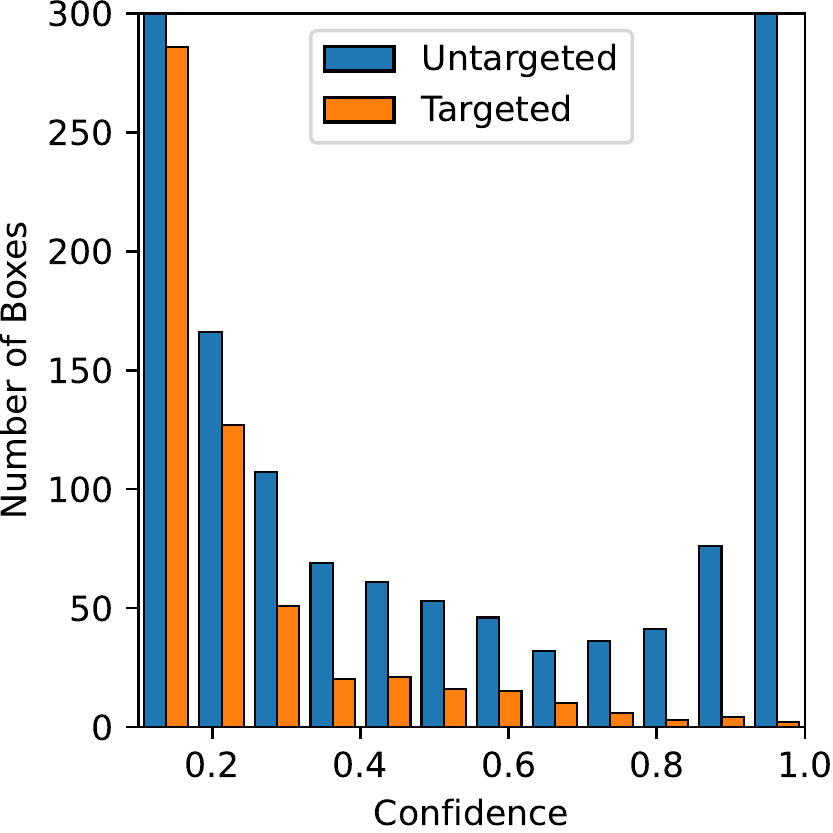}
    }
    \vspace{-0.25in}
    \caption{The distribution of TP and FP bounding boxes under different targeted and untargeted adversarial attacks. ``TP'' represents True Positive, while ``FP'' denotes False Positive.}
    \vspace{-0.1in}
    \label{fig:boxes}
\end{figure}

\revise{As shown in Figure \ref{fig:target}, in terms of mAP, the performance of targeted attacks is weaker than that of untargeted attacks (38.82 v.s. 23.05). However, in terms of FN bounding box numbers, the performance of targeted attacks outperforms untargeted attacks largely (1632 v.s. 1274). These results demonstrate the different adversarial goals for targeted and untargeted attacks, and the targeted attack in the X-ray security inspection scenario might be more meaningful. We conjecture the main reasons for the above observations are as follows: \ding{182} It is more difficult for targeted attacks to reduce the mAP values in general object detection \cite{liu2018dpatch}. \ding{183} As the distribution of all bounding boxes shown in Figure \ref{fig:boxes}, the untargeted attacks produce more false bounding boxes (FP) with high confidence, which could help to reduce the precision of detectors. However, targeted attacks result in fewer FP boxes, and this will help to prevent the detection of prohibited items, but the overall precision will not be too low.}

\revise{\textbf{Unseen prohibited items.} Moreover, we are interested in discovering the potential of \method for attacks on other prohibited items with unseen materials/shapes. In other words, adversarial objects are first generated on specific types of prohibited items, and then we use them to directly attack other unseen prohibited item types without re-training. Here, we first conduct experiments in the digital world, where we train a group of adversarial objects using DOAM on HiXray (unseen prohibited items: lighter, liquids, \emph{etc}) and then directly test them against another DOAM on OPIXray (prohibited items are knives). Overall, our attack achieves 29.40 mAP which is slightly lower than the original \method attack (23.05 mAP) that is trained on OPIXray. We then verify this in the physical world, where we place the adversarial objects around their unseen prohibited items, and we achieve 45.52 mAP, which is also slightly lower than the original \method attack (33.16 mAP). The above results indicate that our attack could still work for other unseen materials/shapes without re-training.}

\section{Countermeasures against \method}
In this section, we propose three possible defenses and evaluate our \method against them. 

\textbf{Data Augmentation.}
Data augmentation has been identified as a popular approach for improving model robustness \cite{rebuffi2021fixing}. In light of this, we introduce the data augmentation strategy as the first countermeasure to mitigate our adversarial attacks. Given the special feature space of X-ray image recognition (\ie, limited colors and textures), we believe that introducing additional adversarial-object-like patches might be beneficial for improving the robustness of the X-ray prohibited item detection models. Specifically, for each image, we randomly add 1-4 blue or orange patches and mix the clean examples with the additional examples during training using a ratio of $1:1$. The results can be found in Table \ref{tab:augmentation}. Here, ``V+C'' denotes that the detector is trained \textbf{without} additional examples and tested \textbf{on} clean examples, ``V+A'' denotes that the detector is trained \textbf{without} additional examples and tested \textbf{on} adversarial examples, ``D+C'' denotes that the detector is trained \textbf{with} additional examples and tested \textbf{on} clean examples and ``D+A'' denotes that the detector is trained \textbf{with} additional examples and tested \textbf{on} adversarial examples. It can be observed that the data augmentation can to a certain extent effectively defend the proposed $\mathcal{X}-$adv method for X-ray prohibited item detection.

\begin{table}[!t]

\caption{Countermeasure studies. (a) ``V'' and ``D'' denote vanilla training or data augmentation; ``C'' and ``A'' refer to testing on clean or adversarial examples; (b) We first generate adversarial examples by meshAdv on DOAM/LIM and train the classifier; we then test the detection performance on \method generated on DOAM. ACC denotes classification accuracy, and AUC is the area under the ROC curve; (c) We adversarially train a prohibited item detector using RobustDet.}
\vspace{-0.1in}

\begin{center}
\subtable[Data augmentation]{
\label{tab:augmentation}
\footnotesize
\setlength{\tabcolsep}{3.0mm}{
\begin{tabular}{@{}ccccccc@{}}
\toprule
\multirow{2}{*}{Setting} & \multirow{2}{*}{mAP} & \multicolumn{5}{c}{Categories}        \\ \cmidrule(l){3-7} 
                         &                      & FO    & ST    & SC    & UT    & MU    \\ \midrule
V+C                      & 74.06                & 78.75 & 40.90  & 95.66 & 73.56 & 81.42 \\
V+A                      & 23.05                & 18.40 & 4.05  & 64.80 & 18.57 & 9.45 \\
D+C                      & 73.94                & 79.44 & 40.52 & 93.82 & 73.40  & 82.54 \\
D+A                      & 46.69                & 49.06 & 17.05 & 81.21 & 39.68 & 46.46 \\ \bottomrule
\end{tabular}
}
}

\subtable[Adversarial detection]{
\label{tab:detection}
\footnotesize
\setlength{\tabcolsep}{5.7mm}{
\begin{tabular}{@{}ccc|cc@{}}
\toprule
\multirow{2}{*}{} & \multicolumn{2}{c|}{DOAM$\rightarrow$DOAM} & \multicolumn{2}{c}{LIM$\rightarrow$DOAM} \\ \cmidrule(l){2-5} 
                        & ACC            & AUC            & ACC              & AUC              \\ \midrule
OPIXray                 & 62.66          & 97.99          & 56.66            & 96.53            \\
HiXray                  & 76.73          & 97.95          & 74.72            & 98.91            \\ \bottomrule
\end{tabular}
}
}

\subtable[Adversarial Training]{
\label{tab:advtraining}
\footnotesize
\setlength{\tabcolsep}{1.5mm}{
\begin{tabular}{cccccccc}
\toprule
\multirow{2}{*}{AT Setting} & \multirow{2}{*}{Attack} & \multirow{2}{*}{mAP} & \multicolumn{5}{c}{Categories}        \\ \cmidrule(l){4-8} 
                            &                         &                      & FO    & ST    & SC    & UT    & MU    \\ \midrule
\multirow{2}{*}{PGD}        & Clean                   & 73.74                & 77.06 & 37.86 & 94.39 & 72.78 & 86.61 \\
                            & \method                 & 22.09                & 20.19 & 1.36  & 66.17 & 17.39 & 5.32  \\ \midrule
\multirow{2}{*}{\revise{\method}} & \revise{Clean}   & \revise{73.49}            & \revise{78.21}  & \revise{40.77}  & \revise{93.23}  & \revise{73.58} & \revise{81.64}   \\
                            & \revise{\method}        & \revise{53.47}      & \revise{55.82} & \revise{20.26} & \revise{84.43} & \revise{49.02}   & \revise{57.82} \\ \bottomrule
\end{tabular}
}
}

\end{center}
\vspace{-0.3in}

\end{table}

\textbf{Adversarial Detection.}
Another prevailing approach to improving model robustness is adversarial detection. Rather than correctly detecting the target item under the adversarial scenario, adversarial detection aims to detect the existence of adversarial examples \cite{feinman2017detecting, carlini2017adversarial, ma2018characterizing, deng2021libre}. Here, we build a neural classifier capable of distinguishing images containing adversarial objects from clean X-ray images. Specifically, we use a ResNet50 model as a classifier and trained on adversarial examples generated by meshAdv on different models (\ie, DOAM and LIM). We then test the classifier on adversarial examples generated by \method on a different DOAM model. The training set and test set of the classifier do not overlap. The results in Table \ref{tab:detection} indicate that the adversarial examples generated by different methods are quite different and the neural classifier fails to generalize.

\textbf{Adversarial Training.} We choose adversarial training (AT) as the last countermeasure for $\mathcal{X}$-\emph{Adv}. Although AT for image classification has been widely studied \cite{madry2017towards,liu2019training}, only some preliminary studies have been devoted to object detection \cite{zhang2019towards, chen2021class, dong2022adversarially}. Here, we adopt RobustDet \cite{dong2022adversarially} as the AT method and use an SSD detector with a backbone of VGG-16. \revise{Specifically, we adversarially train two detectors using adversarial examples generated by (1) PGD attacks or (2) our $\mathcal{X}$-\emph{Adv}. Note that the generated adversarial objects by \method during AT are different from those for testing. From Table \ref{tab:advtraining}, we could observe that \ding{182} AT trained on PGD attacks show limited defense against $\mathcal{X}$-\emph{Adv} mainly due to the differences between perturbations and patch/object attacks; \ding{183} AT trained on \method could mitigate our $\mathcal{X}$-\emph{Adv} attacks to a certain extent; and \ding{184} compared to classification, it is still comparatively difficult to adopt AT on object detection tasks.}

\revise{\textbf{Summary and Discussion.} 
Despite facing significant challenges launched by our attack, the proposed defenses could still mitigate its negative influence to some extent. Specifically, for the strongest countermeasure (AT trained on our $\mathcal{X}$-\emph{Adv}), we could significantly improve the model robustness and achieve 53.47 mAP against \method attacks. Meanwhile, we could observe that the proposed countermeasures have rather high practical feasibility for defenders mainly due to \ding{182} the adversarial-object-like patches of data augmentation and adversarial attacks of adversarial detection can be easily generated and obtained for model training; \ding{183} \method adversarial objects of adversarial training can be generated either based on our open-sourced codes or the provided adversarial objects in the XAD dataset. For the physical-world implementation, defenders could simply employ a 3D printer to print out the 3D objects based on our guidelines. Moreover, defenders could combine adversarial detection with an AT model, which could further mitigate the negative impacts of $\mathcal{X}$-\emph{Adv} attacks. \emph{The feasibility of data augmentation and $\mathcal{X}$-Adv AT are verified under the physical-world setting in Appendix \ref{result}.}}

\section{Physical-world X-ray Attack Dataset}
\label{sec:xad}
A dataset is significantly beneficial for boosting research, especially for areas where professional benchmarks are lacking or the data collection is expensive. As we have observed in Section \ref{sec:phy}, physical-world X-ray detectors are vulnerable to our attacks. Therefore, \final{we further present a physical-world X-ray inspection security robustness evaluation dataset to promote the design of robust X-ray prohibited item detectors.}

\subsection{Construction Details}
We first introduce the construction process of our Physical-world X-ray Attack Dataset (XAD), including the data collection, category selection, and quality control.

\textbf{Data collection}.
We exploit one advanced X-ray security inspection machine, AT150180B, to generate the X-ray images in our dataset. We first randomly place the objects in the plastic/fabric box to mimic the similar environment in the real-world scenario; we then send these boxes through the security inspection machine, after which the machine outputs the X-ray images. To prevent privacy leakage, all images are collected legally: items are borrowed from our staff members and students, and do not contain personal information.

\textbf{Category selection}. As shown in Figure \ref{fig:illus_xad}, we select 4 categories of prohibited items (cutters, scissors, folding knives, straight knives, and utility knives) that frequently appear in daily life. The 4 categories of cutters have different shapes and scales, which meets the category selection diversity requirements. The sufficient numbers of instances can provide a more credible evaluation for various models.

\textbf{Quality control procedure}. We followed a similar annotation quality control procedure to the famous vision dataset Pascal VOC \cite{everingham2010pascal}. All annotators followed the same annotation guidelines, including what to annotate, how to annotate bounding, how to treat occlusion, \etc{} Moreover, to ensure the accuracy of annotation, we divided the annotators into 3 groups. All images were randomly assigned to 2 out of the 3 groups for annotation, after which a final group was specially organized for confirmation.


\subsection{Data Properties}

\begin{figure}[!t]
	\begin{center}
 \vspace{-0.1in}
		\includegraphics[width=1.0\linewidth]{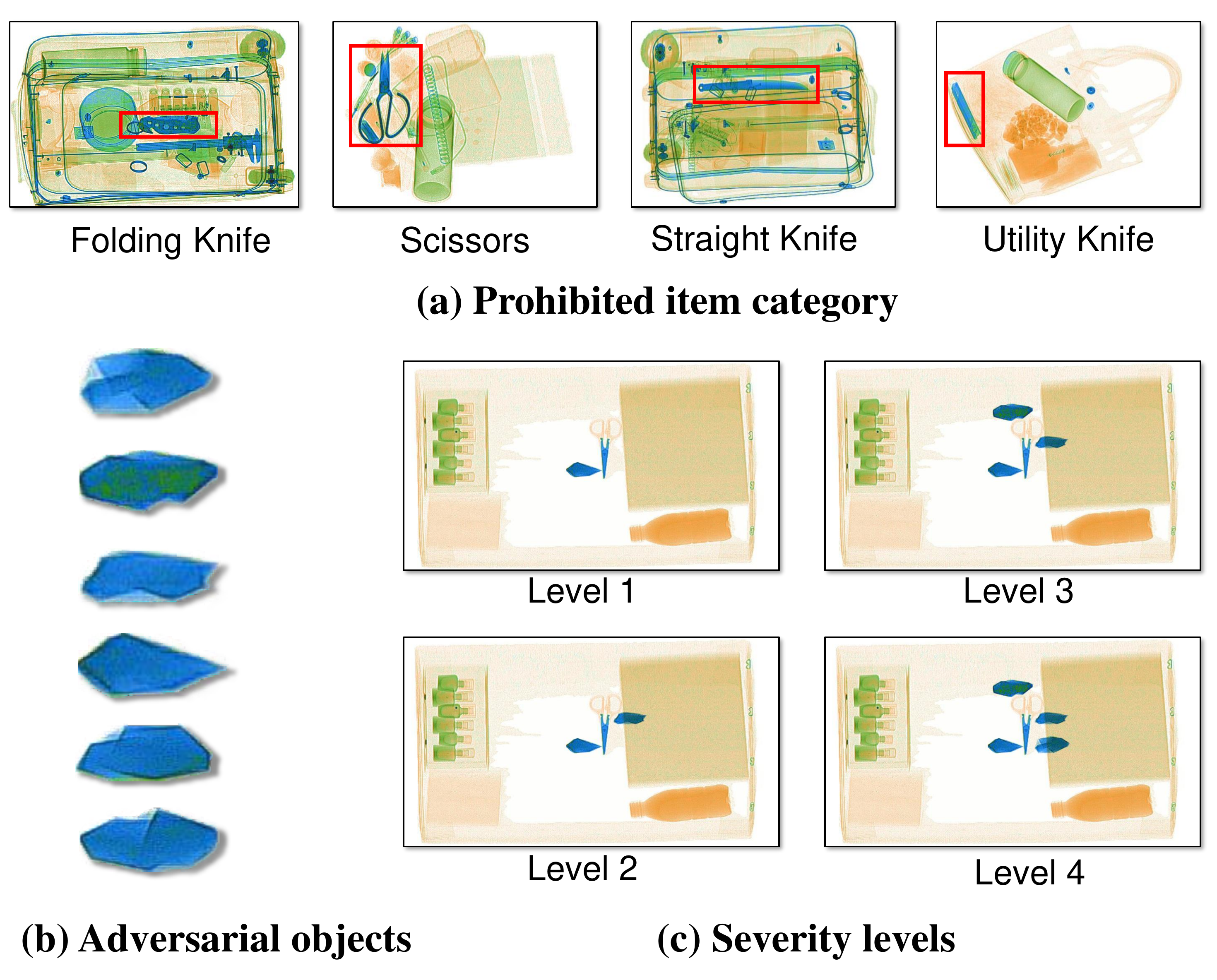}
	\end{center}
	\vspace{-0.2in}
	\caption{Illustration of the images in our XAD. (a) illustrates the prohibited item categories; (b) denotes the X-ray images of physical-world adversarial objects; (c) denotes the different severity levels in the testing set.}
	\label{fig:illus_xad}
	\vspace{-0.1in}
\end{figure}

\quad\textbf{Subset division.} \revise{Our XAD contains two subsets, \ie, a training set with clean X-ray images and a testing set with physical-world adversarial attacks. The 4,537 images in the training set simulate the real-world scenario to help models achieve satisfactory generalization performance. For the testing set, we follow \cite{hendrycks2018benchmarking} and generate adversarial images from 210 clean X-ray images with 4 different severity levels (0 to 4, where ``0'' denotes clean images). Specifically, for each item layout, we first place all the items in the box and X-ray scanned them to obtain a clean image; we then place 1$\sim$4 adversarial metals in the box respectively and scanned them to collect 4 versions of adversarial images with 4 severity levels. Thus, our testing set contains 1,050 samples which are all scanned from a real X-ray machine. See Fig \ref{fig:illus_xad} for samples.}

\textbf{Category distribution.} Our XAD dataset contains 4,537 images and 4 categories of 4,830 instances with bounding-box annotations of prohibited cutters.

\textbf{Color Information}. The colors of objects under X-ray are determined by their chemical composition, mainly reflected in the material, which is introduced in Table \ref{tab:xad_color}.

\textbf{Instances per image.} In the training set of our XAD, each image contains at least one prohibited object. In particular, the image numbers containing 1, 2, 3, and 4 prohibited objects are 4069, 234, 23, and 1, respectively.

\begin{table}[!t]
\vspace{-0.2in}
\caption{Detailed data properties of our XAD dataset.}
\label{table-para_items}
	\begin{center}
	
	\vspace{-0.2in}
	\subtable[Quality distribution]{
	\label{tab:xad_distribution}
		\setlength{\tabcolsep}{0.9mm}{
			\small
			\begin{tabular}{lccccc}
				\toprule
				\textbf{Category} & \textbf{Scissor} & \textbf{Folding knife} & \textbf{Straight knife} & \textbf{Utility knife} \\
				\midrule
				\footnotesize{Training} & \footnotesize{1,048} & \footnotesize{1,300} & \footnotesize{1,300} & \footnotesize{926}  \\
				\footnotesize{Testing} & \footnotesize{54} & \footnotesize{54} & \footnotesize{52} & \footnotesize{50} \\
				\midrule
				\footnotesize{Total} & \footnotesize{2,002} & \footnotesize{13,54} & \footnotesize{1,352} & \footnotesize{976} \\
				\bottomrule
			\end{tabular}
		}
		}
		
	\subtable[Object materials and X-ray image colors]{
	\label{tab:xad_color}
	\small
	\setlength{\tabcolsep}{3.9mm}
	\begin{tabular}{ccllllc}
		\toprule
		\textbf{Colors}  & \multicolumn{5}{c}{\textbf{Materials}}             & \textbf{Typical examples} \\ \midrule
		Orange & \multicolumn{5}{c}{Organic Substances}   & Plastics, Clothes   \\ 
		Blue   & \multicolumn{5}{c}{Inorganic Substances} & Irons, Coppers     \\ 
		Green  & \multicolumn{5}{c}{Mixtures}             & Edge of phones   \\ \bottomrule
	\end{tabular}
		
	}

	\end{center}
	
	\vspace{-0.15in}
\end{table}

\begin{table}[!t]
\vspace{-0.2in}
\caption{Results on different levels of XAD.}
\label{tab:XAD_results}
    \begin{center}
    \vspace{-0.2in}
	
    \footnotesize
    \setlength{\tabcolsep}{4.5mm}{
    \begin{tabular}{@{}cccccc@{}}
    \toprule
    \multirow{2}{*}{Setting} & \multirow{2}{*}{mAP} & \multicolumn{4}{c}{Categories} \\ \cmidrule(l){3-6} 
                             &                      & SC     & FO    & ST    & UT    \\ \midrule
    Level 0                  & 91.74                & 96.29  & 86.98 & 84.86 & 98.84 \\
    Level 1                  & 72.98                & 79.25  & 61.32 & 69.30 & 82.04 \\
    Level 2                  & 50.10                & 66.47  & 33.79 & 60.84 & 39.29 \\
    Level 3                  & 30.83                & 55.76  & 18.59 & 41.15 & 7.82  \\
    Level 4                  & 27.50                & 53.63  & 15.19 & 35.17 & 6.00  \\ \bottomrule
    \end{tabular}
    }
    \end{center}
	\vspace{-0.2in}

\end{table}

\subsection{Preliminary Experiments on XAD}

After introducing our XAD, we further conduct experiments on XAD to demonstrate the difficulties and practicability of maintaining robust detectors. Specifically, we use the DOAM model for detection. We train the model on the training set of XAD and then evaluate it on the test set. The implementation details are the same as our main experiment. From Table \ref{tab:XAD_results}, we can make several observations: \ding{182} The detector shows weak performance on our XAD dataset with the model's performance on prohibited item recognition reducing by as much as 60\% in terms of mAP. \ding{183} Increasing the number of adversarial objects improves the attack and therefore increases the recognition difficulties in this scenario. We encourage researchers to design stronger training strategies or defense modules and evaluate their robustness on this benchmark.

\section{Conclusion and Future Work}

This paper takes the first step to study physical-world adversarial attacks for X-ray prohibited item detection. Specifically, we propose \method{}, which generates physically realizable adversarial objects to circumvent the color fading and complex occlusion problems in this scenario. Although the results presented here are promising, there are several research directions that we are interested in exploring in the future. \ding{182} We hope \method can be used as a tool to better debug and understand the nature of object detectors' robustness. \ding{183} We would like to generate attacks in other soft materials which are more stealthy. \revise{\ding{184} We are interested in attacks against more types of prohibited items.} \ding{185} Our \method can be regarded as a general attacking framework for visually constrained scenarios. In this paper, we focus on attacking X-ray inspection scenarios; we will further extend our attacks to other complex scenarios. 

\section{Ethics Statement}
As an effective way to discover safety problems, adversarial attacks will encourage researchers to pay more attention to model robustness. Based on this, this paper proposes \method to attack X-ray prohibited item detectors. Our large-scale experiments demonstrated that existing X-ray prohibited item detection models (even commercial systems) are not infallible and can still be easily deceived. All experiments are conducted on public-available datasets, and all images for physical-world attacks are collected legally from our staff and students without personal information under their grants. 

\revise{To mitigate potential real-world impacts of the attacks, this paper \ding{182} proposes three countermeasures and discusses their practical feasibility for defending \method; \ding{183} \final{presents the physical-world X-ray attack dataset XAD} to promote the design and re-training of stronger detectors; and \ding{184} disclose the results, countermeasures, and resources to two relevant X-ray security inspection service providers and a stakeholder user at the airport checkpoint. Based on our easy-to-use countermeasures, we help them to recognize this critical security issue and move the first step to improve the robustness of their detection backend with adversarial training. Moreover, these service providers are also suggested to utilize the white-box \method to help reveal the vulnerabilities of their detectors and further design stronger models. Despite the threats identified in this paper, we should note that a real-world adversary would still find it difficult to pass X-ray security inspection systems carrying prohibited items without detection, as human inspectors still help with checking luggage. We thus further suggested the airport checkpoint pay attention to the employment of human inspectors, which can relieve the concerns on the potential negative abuses of $\mathcal{X}$-\emph{Adv}.} 

\section*{Acknowledgement}

The authors would like to thank the shepherd and all the anonymous reviewers for their tremendous efforts and insightful comments during reviewing. This work was supported by the National Key Research and Development Plan of China (2021ZD0110601), the National Natural Science Foundation of China (No. 62022009 and 62206009), the State Key Laboratory of Software Development Environment (SKLSDE-2022ZX-23), and the grants from SenseTime.

\bibliographystyle{plain}
\bibliography{main}
\appendix
\counterwithin{table}{section}
\counterwithin{figure}{section}
\section*{Appendix}
\section{Implementation Details}

\subsection{Pseudo Code of \method Algorithm}
\label{pseudocode}

\begin{algorithm}[!htb]
    \caption{\method Algorithm}
    \label{alg1}
    \begin{algorithmic}[1]
    \renewcommand{\algorithmicrequire}{\textbf{Input:}}
    \renewcommand{\algorithmicensure}{\textbf{Output:}}
    \REQUIRE X-ray image $\mathbf{I}$, label $\mathbf{y}$ and bounding box $\mathbf{b}$ of the prohibited item, initial state of meshes $\mathbf{x}_{ori}$, maximum location reinforcement iterations $N_\mathbf{C}$, and maximum shape polishment iterations $N_\mathbf{P}$.
    \ENSURE Adversarial example $\mathbf{I}_{adv}$, adversarial meshes $\mathbf{x}_{adv}$.
    \STATE Initialize policy network $\pi$.
    \FOR {$i$ \textbf{in} $N_\mathbf{C}$}
        \STATE Sample a location $\mathbf{C}$ in $\pi(\mathbf{I}; \mathbf{w})$.
        \STATE Calculate reward $G$ according to Eqn. \ref{eqn:reward}.
        \STATE Update policy network $\pi$ according to Eqn. \ref{eqn:reinforce}.
    \ENDFOR
    \STATE Initialize $\mathbf{x}_{adv}$ as $\mathbf{x}_{ori}$.
    \FOR {$i$ \textbf{in} $N_\mathbf{P}$}
        \STATE Generate $\mathbf{I}_{adv} = \mathbf{I} \odot g_m(d(\mathbf{x}_{adv}^{\mathbf{P, C}}))$.
        \STATE Calculate cost function $\mathcal{L}$ according to Eqn. \ref{eqn:totalloss}.
        \STATE Update $\mathbf{x}_{adv}$ by Adam optimizer.
    \ENDFOR
    \end{algorithmic}
\end{algorithm}

\subsection{X-ray Converter}
\label{sec:converter}
\begin{figure}[!t]
	\begin{center}
		\includegraphics[width=1.0\linewidth]{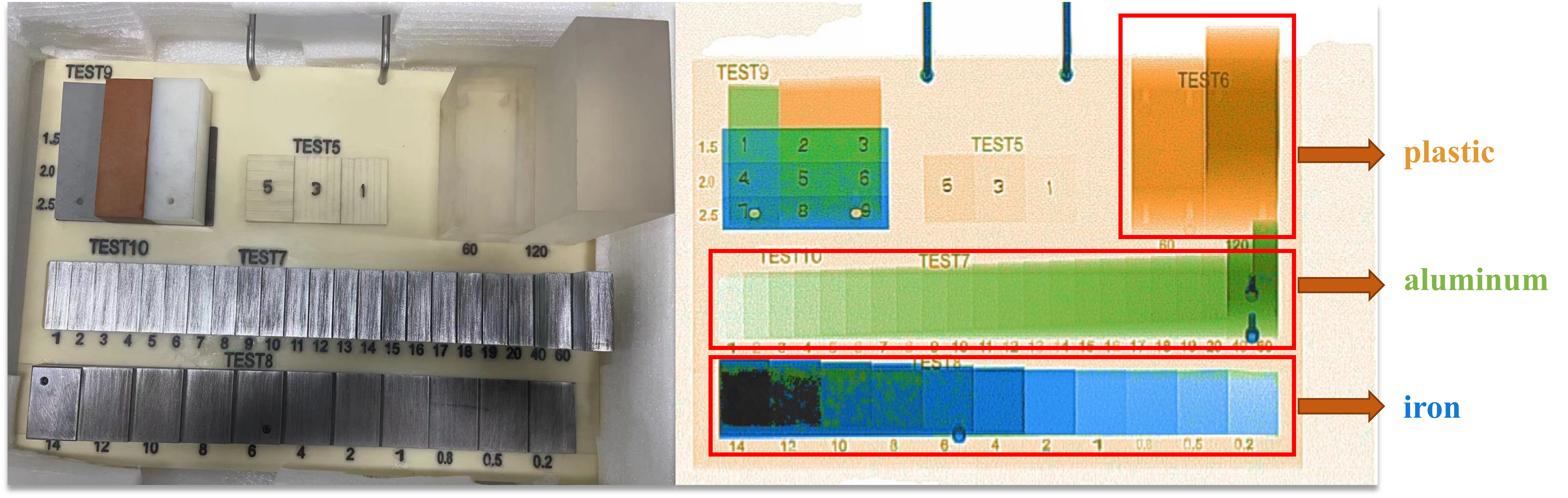}
	\end{center}
	\vspace{-0.15in}
	\caption{Physical-world photos and X-ray images of scanned objects with different materials and thicknesses. The number under the objects denotes the thickness of the object in millimeters.}
        \label{fig:plate}
	\vspace{-0.1in}
\end{figure}

\begin{figure}[!t]
	\begin{center}
		\includegraphics[width=0.7\linewidth]{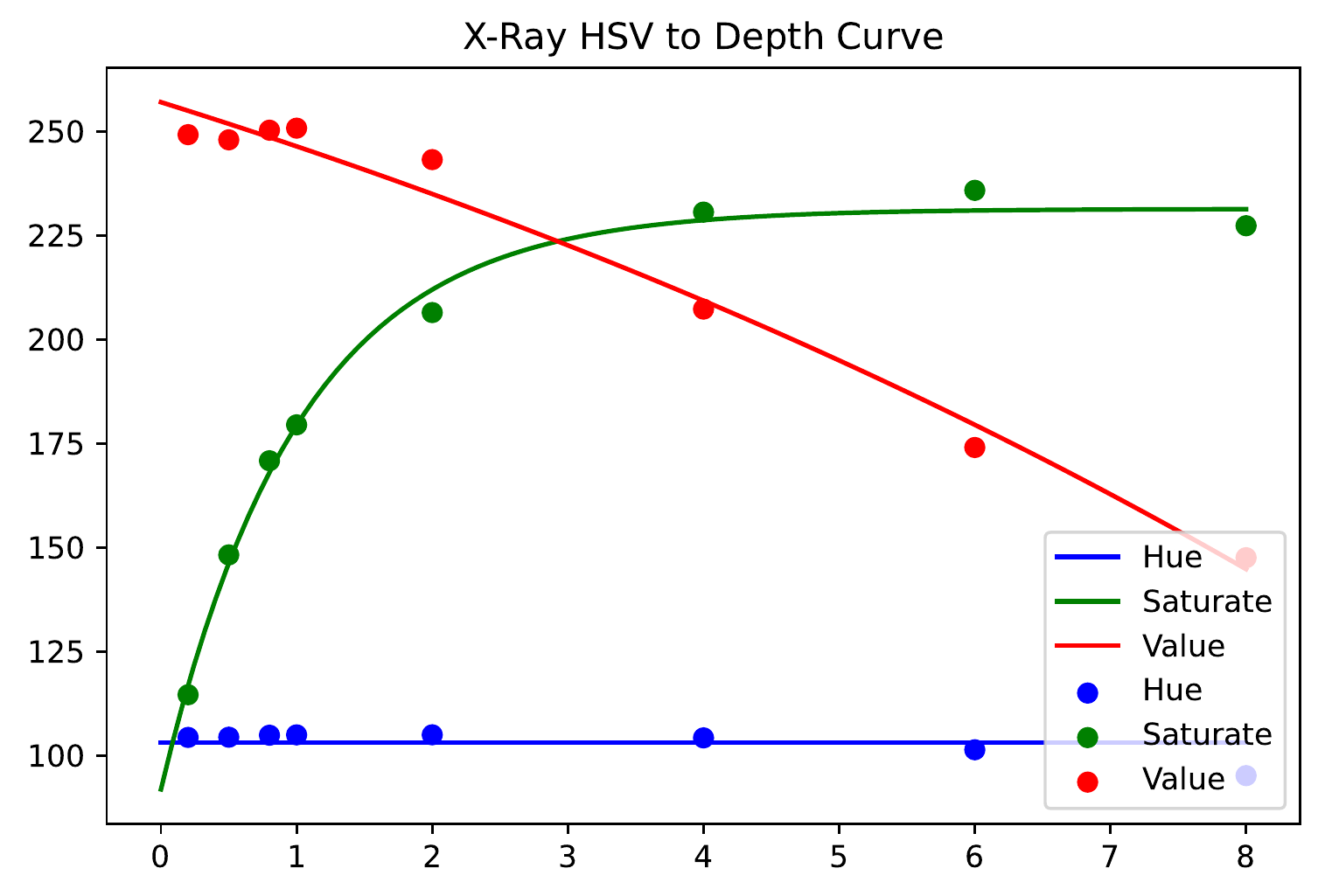}
	\end{center}
	\vspace{-0.2in}
	\caption{Regression curves of depths to HSV values for \texttt{iron} material. We sample 8 different thicknesses from 0.2mm to 8mm.}
        \label{fig:hsv_curve}
	\vspace{-0.2in}
\end{figure}

To obtain the coefficient of the X-ray converter, we have scanned a group of objects with different types of materials and thicknesses. The photo of the scanned objects and their X-ray images is provided in Figure \ref{fig:plate}. For every thickness and material, we sample an area of $20\times20$ pixels and calculate the average pixel value as the color for this specific thickness and material. We use Eqn. \ref{eqn:gmd} as the conversion function, fitting the coefficient $a, b, c$ to the depth and color pairs. The illustration of regression curves for \texttt{iron} is shown in Figure \ref{fig:hsv_curve}. 

The goodness of fit $R^2$ for Hue, Saturate, and Value is 0 (Hue is a constant value), 0.993, and 0.984, which demonstrates that the proposed conversion function fits well.

\subsection{Detailed Settings of Experiments}
\label{sec:moresetting}

\textbf{Training of target models.} All models in our paper (\ie, SSD, DOAM, LIM, and Faster R-CNN) use VGG-16 as backbones. We use pre-trained weights on the VOC0712 dataset and fine-tune our model on them, which is applied by previous works \cite{wei2020occluded, tao2021towards} to reduce the training time. We use the SGD optimizer with momentum 0.9 and weight decay $5 \times 10^{-4}$. For SSD, DOAM, and LIM, we train them with batch size 24 and a learning rate of 0.0001. For Faster R-CNN, we train them with batch size 1 and a learning rate of 0.001. All models are trained with a maximum of 100 epochs, and we select the checkpoints with the highest mAP as our target models.

\textbf{\method.} We adopt an Adam optimizer with a learning rate of 0.1 and a maximum of 24 iterations to optimize the adversarial loss. To accelerate the speed of attacks, we set the batch size of the attack as 10, which means that every 10 images share the same group of adversarial objects. Experimental results have proven the viability of this approach. The initial shape of the 3D object is a sphere with 26 vertices and 48 faces. During optimization, the coordinates of vertices are updated with the guidance of the gradients, while the adjacent relation of vertices remains unchanged. The coefficient of location variance $\alpha$ is 0.05; the coefficient of perceptual loss $\beta$ is 0.1 in SSD, DOAM, and LIM, and 0.01 in Faster R-CNN. We train the REINFORCE policy for 200 iterations for every batch. For the available attack area $\mathbb{C}$, we define $(x_{min}, y_{min}), (x_{max}, y_{max})$ as the coordinates of a ground truth box, and $w, h$ is the width and height of the box. Adversarial objects should have no overlap with the center area of the ground truth box, \ie, the area of $(x_{min}+0.25w, y_{min}+0.25h), (x_{max}-0.25w, y_{max}-0.25h)$. 

\textbf{Vanilla adversarial objects.} We simply set the optimizing epoch to 0 to get vanilla spheres as adversarial objects.

\textbf{MeshAdv \cite{xiao2019meshadv}.} We apply the same loss with \method to perform MeshAdv attacks. MeshAdv does not have the X-ray converter and location reinforcement; thus we fix the color of the converter and set the attack location as the four corners of 
the available attack areas.

\textbf{Adversarial Patch \cite{brown2017adversarial}.} We apply the first term of \method loss (\ie, $\mathcal{L}_{adv}$) for Adversarial Patch attacks. The initial shape of the adversarial patch is a $20 \times 20$ depth image filled with 0. To utilize a 2D patch, the gradient is calculated and updated on the depth image rather than a 3D object. We set the attack location as the four corners of available attack areas.


\subsection{Time Consumption of \method}

In Table \ref{table:timecost}, we show the time consumption of our \method attack on OPIXray dataset for the SSD detector. All of our experiments are conducted on 1 GPU of NVIDIA RTX 3080Ti and 8 CPU cores of Intel Xeon Gold 6148 @2.40GHz. The results demonstrate that our \method generates adversarial attacks with reasonable time consumption.

Also, we should note that the location reinforcement process takes the most time during the attack generation, and we will accelerate this process in future work.

\begin{table}[!t]
\begin{center}

\caption{Time consumption of our \method on OPIXray dataset for the SSD detector. We perform 24 iterations of shape polishment and 200 iterations of location reinforcement, which is consistent with other experiments. We show the time consumption of one iteration and one batch.}
\label{table:timecost}
\vspace{0.1in}
\footnotesize
\begin{tabular}{@{}ccc@{}}
\toprule
Time Cost (s) & Shape Polishment & Location Reinforcement \\ \midrule
Iteration     & 0.51             & 0.28                   \\
Batch         & 12.15            & 55.93                  \\ \bottomrule
\end{tabular}
\end{center}

\vspace{-0.3in}

\end{table}

\section{Additional Experimental Results}
\label{result}

\begin{table}[!t]
\begin{center}
\caption{Performance comparison between different materials on HiXray dataset.}
\vspace{-0.1in}
\label{table:material_hix}

\footnotesize
\resizebox{\linewidth}{!}{
\begin{tabular}{@{}cccccccccc@{}}
\toprule
\multirow{2}{*}{Setting} & \multirow{2}{*}{mAP} & \multicolumn{8}{c}{Categories}                                                                                                    \\ \cmidrule(l){3-10} 
                         &                      & PO1           & PO2           & WA             & LA             & MP             & TA             & CO            & NL            \\ \midrule
Plastic                  & 55.22                & 43.10         & 39.12         & 64.01          & 96.48          & 91.79          & 77.15          & 30.09         & \textbf{0.00} \\
Aluminum                 & 43.79                & 23.11         & 24.27         & \textbf{51.41} & \textbf{94.44} & 83.11          & 72.67          & \textbf{1.33} & \textbf{0.00} \\
Iron                     & \textbf{38.96}       & \textbf{5.21} & \textbf{3.33} & 63.00          & 95.49          & \textbf{77.38} & \textbf{63.05} & 4.22          & \textbf{0.00} \\ \bottomrule
\end{tabular}

}

\end{center}

\vspace{-0.3in}

\end{table}

Different from OPIXray dataset, images in HiXray dataset may contain more than one prohibited item. We only attack images with only one prohibited item (about 3,227 images) for simplicity. The performance of different materials on the HiXray dataset is shown in Table \ref{table:material_hix}; \revise{the physical-world results on LIM are shown in Table \ref{table:physical_appendix}; the countermeasure validation in the physical world is shown in Table \ref{tab:countermeasure_physical};}the black-box attack results on OPIXray and HiXray are shown in Table \ref{table:blackbox}; the white-box attack results on HiXray are shown in Table \ref{table:digital_hix}. These results demonstrate the effectiveness.

\begin{table}[!t]
\begin{center}
\caption{Additional results of physical-world attack experiments on LIM.}
\label{table:physical_appendix}

\footnotesize
\setlength{\tabcolsep}{3.5mm}{
\begin{tabular}{@{}cccccc@{}}
\toprule
\multirow{2}{*}{Setting} & \multirow{2}{*}{mAP} & \multicolumn{4}{c}{Categories} \\ \cmidrule(l){3-6} 
                         &                      & SC    & FO    & ST     & UT    \\ \midrule
Clean                    & 96.20                & 98.85  & 99.55 & 95.58 & 90.82 \\
Digital attack           & 29.56                & 76.69  & 4.08  & 31.06 & 6.41  \\
Physical best            & 29.46                & 73.88  & 8.80  & 24.09 & 11.04 \\
Physical change          & 51.38                & 75.30  & 47.94 & 41.46 & 40.81 \\
\revise{Physical random} & \revise{77.26}       & \revise{88.56} & \revise{86.33} & \revise{81.60}  & \revise{52.53}

\\
\bottomrule
\end{tabular}
}
\end{center}
\vspace{-0.1in}

\end{table}

\begin{table}[!t]

\caption{\revise{Countermeasure studies in the physical world. We train the defended models on the XAD dataset in the digital world, and evaluate their performance by collecting real images in the physical-world scenario.}}
\label{tab:countermeasure_physical}
\vspace{-0.1in}

\begin{center}
\subtable[\revise{Data augmentation}]{
\label{tab:augmentation}
\footnotesize
\setlength{\tabcolsep}{4.3mm}{
\begin{tabular}{@{}cccccc@{}}
\toprule
\multirow{2}{*}{\revise{Setting}} & \multirow{2}{*}{\revise{mAP}} & \multicolumn{4}{c}{\revise{Categories}} \\ \cmidrule(l){3-6} 
                         &                      & \revise{SC} & \revise{FO} & \revise{ST} & \revise{UT}  \\ \midrule
\revise{V+C} & \revise{91.35} & \revise{84.17} & \revise{98.05} & \revise{100.00} & \revise{83.18} \\
\revise{V+A} & \revise{33.16} & \revise{66.33} & \revise{18.35} & \revise{44.48} & \revise{3.46} \\
\revise{D+C} & \revise{90.57} & \revise{85.76} & \revise{97.14} & \revise{98.42} & \revise{80.98} \\
\revise{D+A} & \revise{51.16} & \revise{63.15} & \revise{31.79} & \revise{72.27} & \revise{37.42} \\ \bottomrule
\end{tabular}
}
}

\subtable[\revise{Adversarial Training}]{
\label{tab:advtraining}
\footnotesize
\setlength{\tabcolsep}{2.2mm}{
\begin{tabular}{cccccccc}
\toprule
\multirow{2}{*}{\revise{AT Setting}} & \multirow{2}{*}{\revise{Attack}} & \multirow{2}{*}{\revise{mAP}} & \multicolumn{4}{c}{\revise{Categories}}        \\ \cmidrule(l){4-7} 
                            &                         &                      & \revise{SC}    & \revise{FO}    & \revise{ST}    & \revise{UT}     \\ \midrule
\multirow{2}{*}{\revise{\method}} & \revise{Clean}   & \revise{91.12}  & \revise{95.03}  & \revise{98.85}  & \revise{100.00} & \revise{70.59} \\
                            & \revise{\method}        & \revise{59.15}      & \revise{89.18} & \revise{49.64} & \revise{80.25} & \revise{17.54} \\ \bottomrule
\end{tabular}
}
}

\end{center}
\vspace{-0.3in}

\end{table}

\begin{table}[!t]
\begin{center}
\caption{Mean average precision of digital-world black-box attacks on OPIXray and HiXray datasets. The adversarial examples are generated from the source model and evaluated on the target model. \textbf{Bold} results denote the best performance of attacking in each column.}
\label{table:blackbox}

\subtable[OPIXray]{
\footnotesize
\setlength{\tabcolsep}{3.8mm}{
\begin{tabular}{@{}ccccc@{}}
\toprule
\multirow{2}{*}{Source Model} & \multicolumn{4}{c}{Target Model}                                  \\ \cmidrule(l){2-5} 
                              & SSD            & Faster R-CNN   & DOAM           & LIM            \\ \midrule
SSD                           & \textbf{19.20} & 30.01          & 44.37          & 39.31          \\
Faster R-CNN                  & 45.60          & \textbf{23.33} & 55.12          & 53.67          \\
DOAM                          & 36.46          & 28.69          & \textbf{23.05} & 40.00          \\
LIM                           & 26.90          & 28.37          & 38.59          & \textbf{22.46} \\ \bottomrule
\end{tabular}
}
}

\subtable[HiXray]{
\footnotesize
\setlength{\tabcolsep}{3.8mm}{
\begin{tabular}{@{}ccccc@{}}
\toprule
\multirow{2}{*}{Source Model} & \multicolumn{4}{c}{Target Model}                                  \\ \cmidrule(l){2-5} 
                              & SSD            & Faster R-CNN   & DOAM           & LIM            \\ \midrule
SSD                           & \textbf{33.41} & 41.28          & 45.02          & 40.45          \\
Faster R-CNN                  & 50.32          & \textbf{39.39} & 48.18          & 48.75          \\
DOAM                          & 45.94          & 42.09          & \textbf{38.96} & 44.87          \\
LIM                           & 42.27          & 47.94          & 44.99          & \textbf{32.53} \\ \bottomrule
\end{tabular}
}
}
\end{center}

\vspace{-0.1in}

\end{table}

\begin{table}[!t]
\begin{center}
\caption{Digital-world white-box attacking results on HiXray dataset. PO1, PO2, WA, LA, MP, TA, CO, and NL denote ``Portable charger 1 (lithium-ion prismatic cell)'', ``Portable charger 2 (lithium-ion cylindrical cell)'', ``Water'', ``Laptop'', ``Mobile Phone'', ``Tablet'', ``Cosmetic'' and ``Nonmetallic Lighter''.}
\label{table:digital_hix}

\subtable[SSD]{
\footnotesize
\resizebox{\linewidth}{!}{
\begin{tabular}{@{}cccccccccc@{}}
\toprule
\multirow{2}{*}{Setting} & \multirow{2}{*}{mAP} & \multicolumn{8}{c}{Categories}                                                                                                    \\ \cmidrule(l){3-10} 
                         &                      & PO1           & PO2           & WA             & LA             & MP             & TA             & CO            & NL            \\ \midrule
Clean    & 63.06          & 58.76          & 57.61         & 74.24          & 97.55          & 95.18          & 80.66          & 40.47         & 0.02          \\
Vanilla  & 56.94          & 54.33          & 42.65         & 73.99          & 96.81          & 92.43          & 77.87          & 17.46         & 0.00          \\
MeshAdv  & 48.14          & 40.90          & 22.56         & 63.79          & 96.45          & 86.20          & 66.40          & 8.82          & 0.00          \\
AdvPatch & 43.71          & 29.55          & 6.62          & 65.95          & 96.65          & 77.62          & 59.98          & 13.30         & 0.00          \\
\method  & \textbf{33.41} & \textbf{10.96} & \textbf{3.52} & \textbf{46.91} & \textbf{95.11} & \textbf{57.17} & \textbf{52.89} & \textbf{0.67} & \textbf{0.00} \\ \bottomrule
\end{tabular}
}
}

\subtable[Faster R-CNN]{
\footnotesize
\resizebox{\linewidth}{!}{
\begin{tabular}{@{}cccccccccc@{}}
\toprule
\multirow{2}{*}{Setting} & \multirow{2}{*}{mAP} & \multicolumn{8}{c}{Categories}          \\ \cmidrule(l){3-10} 
                         &                      & PO1           & PO2           & WA             & LA             & MP             & TA             & CO            & NL            \\ \midrule
Clean    & 66.91          & 74.06          & 62.51         & 81.19          & 98.43          & 95.60          & 80.66          & 42.86         & \textbf{0.00} \\
Vanilla  & 58.60          & 56.04          & 38.51         & 80.58          & 97.73          & 94.49          & 80.10          & 21.38         & \textbf{0.00} \\
MeshAdv  & 51.12          & 28.53          & 17.84         & 77.00          & 98.20          & 92.64          & 80.19          & 14.54         & \textbf{0.00} \\
AdvPatch & 48.37          & 42.05          & \textbf{1.40} & 77.37          & 97.91          & 82.84          & 74.79          & 10.56         & \textbf{0.00} \\
\method  & \textbf{39.39} & \textbf{21.91} & 4.54          & \textbf{62.73} & \textbf{94.33} & \textbf{70.99} & \textbf{56.98} & \textbf{3.64} & \textbf{0.00} \\ \bottomrule
\end{tabular}

}
}

\subtable[DOAM]{
\footnotesize
\resizebox{\linewidth}{!}{
\begin{tabular}{@{}cccccccccc@{}}
\toprule
\multirow{2}{*}{Setting} & \multirow{2}{*}{mAP} & \multicolumn{8}{c}{Categories}                                                                                                    \\ \cmidrule(l){3-10} 
                         &                      & PO1           & PO2           & WA             & LA             & MP             & TA             & CO            & NL            \\ \midrule
Clean                    & 63.43                & 64.71         & 57.02         & 72.11          & 97.14          & 94.68          & 80.86          & 41.70         & 0.01          \\
Vanilla                  & 54.23     & 44.33     & 30.91     & 74.77     & 96.65     & 92.43     & 80.39     & 14.32     & \textbf{0.00} \\
MeshAdv                  & 45.93                & 12.24         & 10.01         & 73.35          & 96.41          & 89.54          & 75.82          & 10.09         & \textbf{0.00} \\
AdvPatch                 & 42.20                & 16.27         & \textbf{0.31} & 67.50          & 96.02          & 85.05          & 67.20          & 5.26          & \textbf{0.00} \\
\method                  & \textbf{38.96}       & \textbf{5.21} & 3.33          & \textbf{63.00} & \textbf{95.49} & \textbf{77.38} & \textbf{63.05} & \textbf{4.22} & \textbf{0.00} \\ \bottomrule
\end{tabular}
}
}

\subtable[LIM]{
\footnotesize
\resizebox{\linewidth}{!}{
\begin{tabular}{@{}cccccccccc@{}}
\toprule
\multirow{2}{*}{Setting} & \multirow{2}{*}{mAP} & \multicolumn{8}{c}{Categories}                                                                                                    \\ \cmidrule(l){3-10} 
                         &                      & PO1           & PO2           & WA             & LA             & MP             & TA             & CO            & NL            \\ \midrule
Clean                    & 64.84                & 69.89         & 57.85         & 69.98          & 97.89          & 95.10          & 81.71          & 46.24         & 0.08          \\
Vanilla                  & 55.34     & 49.03     & 34.09     & 66.35     & 98.10     & 91.74     & 75.41     & 28.00     & 0.01      \\
MeshAdv                  & 46.51                & 29.23         & 11.38         & 60.59          & 97.90          & 86.77          & 71.13          & 15.09         & \textbf{0.00} \\
AdvPatch                 & 40.75                & 25.65         & \textbf{0.31} & 52.11          & 97.98          & 76.88          & 61.20          & 11.83         & \textbf{0.00} \\
\method                  & \textbf{32.53}       & \textbf{2.98} & 0.70          & \textbf{37.94} & \textbf{97.18} & \textbf{60.61} & \textbf{58.82} & \textbf{1.99} & \textbf{0.00} \\ \bottomrule
\end{tabular}
}
}

\end{center}
\vspace{-0.15in}
\end{table}

\end{document}